\documentclass[10pt, conference, compsocconf]{IEEEtran}
\usepackage{amsmath,epsfig,color,url}
\usepackage{amsfonts}
\usepackage{graphicx}
\usepackage{amssymb}
\usepackage{algorithmic}
\usepackage{algorithm}
\usepackage{tikz}
\usepackage{xcolor}
\usepackage{multirow}
\usetikzlibrary{shapes.misc}
\usetikzlibrary{positioning}

\newcommand{\power}{P_T}
\newcommand{\T}{\mathcal{T}}
\newcommand{\M}{\mathcal{A}}
\newcommand{\X}{\mathcal{X}}
\newcommand{\V}{\mathcal{V}}

\newcommand{\solSpace}{\mathcal{S}}
\newcommand{\bound}{\mathcal{B}}
\newcommand{\E}{\mathcal{E}}
\newcommand{\R}{\mathcal{R}}
\newcommand{\Nout}[2]{ \overrightarrow{ \mathcal{N}_{#1}^{#2} } }
\newcommand{\Nin}[2]{ \overleftarrow{ \mathcal{N}_{#1}^{#2} } }

\newcommand{\paths}{\mathbf{p}}

\newcommand{\Qmatrix}{Q}

\newcommand{\Qe}[4]{\Qmatrix_{#1#2}^{#3#4}}

\newcommand{\minitab}[2][l]{\begin{tabular}{#1}#2\end{tabular}}
\begin{document}
\title{Deriving Pareto-optimal performance bounds for 1 and 2-relay wireless networks}

%\author{No author given}
\author{\IEEEauthorblockN{Qi Wang \IEEEauthorrefmark{1}\IEEEauthorrefmark{3}, Katia Jaffr\`es-Runser\IEEEauthorrefmark{2},
Claire Goursaud \IEEEauthorrefmark{3}
and
Jean-Marie Gorce\IEEEauthorrefmark{3},
}
\IEEEauthorblockA{\IEEEauthorrefmark{1}Institute of Computing Technology, Chinese Academy of Sciences,
Beijing, CHINA\\
Email: wangqi08@ict.ac.cn}
\IEEEauthorblockA{\IEEEauthorrefmark{2}Universit\'e de Toulouse, IRIT / ENSEEIHT,
F-31061, Toulouse, FRANCE\\
Email: katia.jaffres-runser@irit.fr}
\IEEEauthorblockA{\IEEEauthorrefmark{3}Universit\'e de Lyon, INRIA,
INSA-Lyon, CITI, F-69621, FRANCE\\
Email: \{claire.goursaud, jean-marie.gorce\}@insa-lyon.fr}
}

\maketitle

\begin{abstract}
This work addresses the problem of deriving fundamental trade-off bounds for a 1-relay and a 2-relay wireless network when multiple performance criteria are of interest. It proposes a simple MultiObjective (MO) performance evaluation framework composed of a broadcast and interference-limited network model; capacity, delay and energy performance metrics and an associated MO optimization problem. Pareto optimal performance bounds between end-to-end delay and energy for a capacity-achieving network are given for 1-relay and 2-relay topologies and assessed through simulations. Moreover, we also show in this paper that these bounds are tight since they can be reached by simple practical coding strategies performed by the source and the relays. Two different types of network coding strategies are investigated. Practical performance bounds for both strategies are compared to the theoretical upper bound. Results confirm that the proposed upper bound on delay and energy performance is tight and can be reached with the proposed combined source and network coding strategies.
\end{abstract}



\section{Introduction}\label{sec:introduction}
Two main and complementary directions have driven research in wireless ad hoc networking.
The first direction targets the design of efficient distributed protocols at all layers of the protocol stack: physical, medium access control (MAC), routing, and transport layers.
Various techniques in the context of resource allocation (power control \cite{Wu01012002}, scheduling, frequency assignment,... ), coding (source coding \cite{Mackay2005, Luby2002}, network coding \cite{Ahlswede2000}, \cite{Ho2006}), and routing (reactive routing \cite{AODV}, proactive routing \cite{OLSR}, opportunistic routing \cite{Jacquet-JSAC2009}, geographic routing \cite{GARE}... ).
The second research direction targets the derivation of fundamental performance limits of wireless ad hoc networks (cf. \cite{Goldsmith2011} and the references herein). Both directions are clearly related since performance limits can provide insight into proper network design solutions and thus, help improving protocol performance. They provide as well upper bounds against which to compare the performance of existing protocols.

Initial research in both directions has concentrated on deriving upper bounds \cite{Toumpis2003, Luo2010} and protocols maximizing network capacity \cite{OLSR, AODV}. Yet, capacity achieving strategies and related bounds even for some simple network configurations are still to be found \cite{Goldsmith2011, Avestimehr2011}.
With the introduction of new applications (e.g. wireless sensor networks, vehicular networks, etc...), additional metrics and their impact on network capacity have become relevant. New studies on the trade-off between metrics implying energy consumption minimization \cite{Gorce2010, Zhang2009}, end to end delay minimization  \cite{Comaniciu2006,Gorce2010} or reliability maximization \cite{Zhang2009} have started. These trade-offs can be characterized with MultiObjective (MO) bounds. A 2-objective MO bound represents the relationship between two criteria $f_1$ and $f_2$.

As considered by Goldsmith et al. in \cite{Goldsmith2011}, a promising way towards achieving fundamental MO bounds in wireless ad hoc networks is to leverage ``the broadcast features of wireless transmissions through generalized network coding, including cooperation and relaying".
In our previous work \cite{Jaffres-Runser2010-WiOpt, RR-upper}, we have proposed a framework composed of a cross-layer network model and a steady state performance evaluation model capturing capacity, delay and energy metrics. We have formulated an associated MO optimization problem whose resolution provides both the MO bound and MO Pareto-optimal network configurations. This framework has been designed to incorporate broadcast and interference-limited channels and thus, is capable of deriving MO bounds for a \emph{layerless} communication paradigm \cite{Goldsmith2011} that integrates generalized network coding, cooperation and relaying.

The purpose of this paper is to assess the quality of this MO bound through the derivation of a lower achievable MO bound. An achievable MO lower bound can be obtained with any distributed network strategy incorporating relaying, coding or cooperation decision. Our aim is to exhibit MO lower bounds that are as close as possible to our MO upper bound, validating the tightness of our MO bound and the efficiency of the network strategy (which is nothing else than a distributed network protocol).
Proposed lower bounds are achieved using simple source and network coding algorithms. Looking at first for simple transmission and relaying strategies is motivated by their ease of deployment. Focusing on network coding is driven by the fact that it leverages the inherent broadcast nature of wireless propagation, phenomenon that is captured as well in the framework used to derive MO upper bounds.
Investigated network strategies have sources transmitting a random linear fountain code and relays re-combining packet using different simple network coding strategies. Two different network coding strategies are investigated. Practical performance bounds for both strategies are compared to the theoretical bound. Therefore, we focus on 1-relay and 2-relay topologies.
Results clearly demonstrate the tightness of our upper MO bound compared to a combined source and network code.

This paper is organized as follows. Our network model is introduced in Section \ref{sec:networkmodel}. The considered MO optimization problem is presented in Section \ref{sec:problemstatement} and its derivation for 1-relay and 2-relay cases in Section \ref{sec:1relay2relay}. Results are given in Section \ref{sec:results} and coding strategies are discussed in Section \ref{sec:coding}. Finally, Section \ref{sec:conclu} concludes the paper.

%------------------------------------------------------------------------------------------------
%----------------------------- SECTION SYSTEM MODEL -------------------------
%------------------------------------------------------------------------------------------------

%II- System Model
%ÊA- Network topologies (define 1 and 2-relay topologies)
%ÊB- Pareto Bounds and Solution Set (define them)
%ÊC- Protocol and network model (define network model for both topologies : link proba, transmission rate and forwarding probability)
\section{System model}\label{sec:networkmodel}
\begin{figure*}[!]
\begin{center}
  \scalebox{0.4}{
  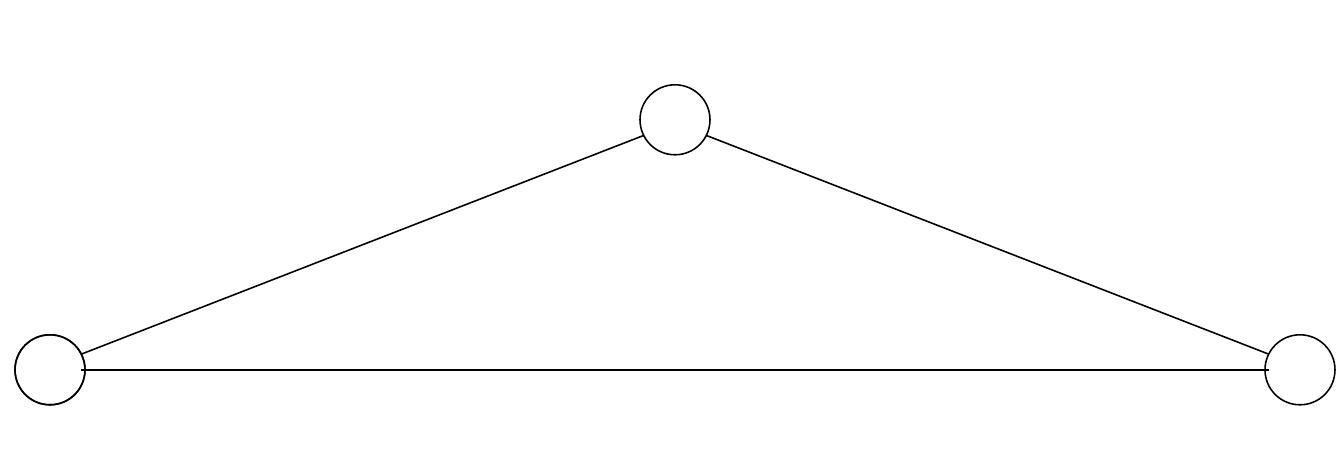 \hskip5cm
  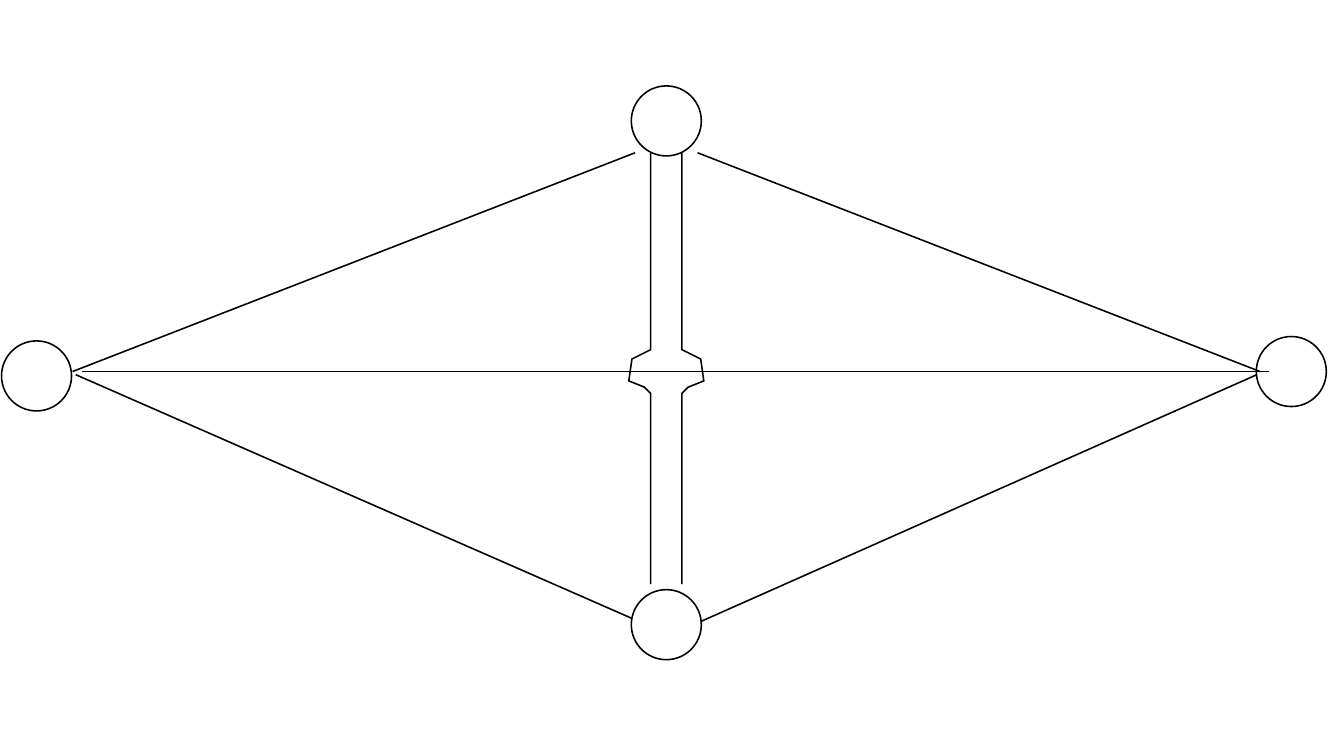}
   \caption{1-relay and 2-relay network topologies}
\label{fig:topologies}
\end{center}
\end{figure*}

The two following topologies of wireless ad hoc networks, illustrated in Fig.~\ref{fig:topologies} are studied in depth in this paper:
\begin{itemize}
\item 1-relay topology
\item 2-relay or diamond topology
\end{itemize}
Next, the framework for our study is introduced.
\subsection{Protocol and network model}

We assume a synchronized wireless ad hoc network where transmissions are time-multiplexed (the synchronization procedure is out of the scope of this paper). A frame of $|\T|$ time slots is repeated indefinitely. One or more packets can be transmitted in a time slot. In the rest of this paper, our examples assume that one packet is being sent in one time slot.
A \emph{time epoch} $s$ is defined as the time needed to transmit one frame of $|\T|$ time slots.

\subsubsection{Wireless channel model}
For any time slot $u \in \T$, there is an interference-limited channel between any two nodes $i$ and $j$ of the network. This channel is modeled by the probability of a packet to be correctly transmitted between $i$ and $j$ in time slot $u$. This probability is referred to as the \emph{channel probability} and denoted $p_{ij}^u$ in the following.

It is computed assuming interference is modeled as an additive noise and for the medium access scheme presented hereafter. Its derivation is based on the distribution of the packet error rates (PER) originating from the statistics of nodes attempting emission in the same time slot. We refer the reader to \cite{Jaffres-Runser2010-WiOpt}\cite{RR-upper} to get the exact derivation of this channel probability.

\begin{figure}[h]
\begin{center}
  \scalebox{0.45}{
  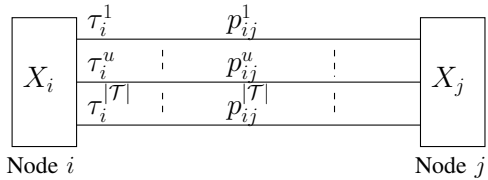}
  \caption{Channel and node model}
\label{fig:networkmodel}
\end{center}
\end{figure}

Each channel is assumed to be in a half-duplex mode, i.e. a node cannot transmit and receive a packet at the same time.

\subsubsection{Network model}
The considered 1-relay and 2-relay wireless networks are modeled by a finite weighted multiple edges complete graph $\mathcal{K_{|\V|}}=\mathcal{(\V,\E)}$ with $\V$ the set of vertices and $\E$ the set of edges. Two vertices are linked by $|\T|$ edges representing orthogonal interference-limited channels as illustrated on Fig.~\ref{fig:networkmodel}.
In this graph, an edge $(i,j,u)$ represents the channel between nodes $i$ and $j$ in time slot $u$. Each edge is assigned a weight of $p_{ij}^u$. If the transmission between $i$ and $j$ on time slot $u$ is not possible, $p_{ij}^u=0$.
For each node $i \in \V$, $\Nout{i}{}$ and $\Nin{i}{}$ are the set of edges leaving from and going into $i$, respectively.

A unique flow with a source $S$ continuously transmitting data to a destination $D$ is defined.
Source and destination nodes do not relay the information.
Multi-hop transmissions are allowed and we model the other nodes as relay nodes $\R = \V-S-D$. We have $N=|\R|$ the number of relays in the network.

As said before, the network is synchronized. Depending on their time slot assignments, source and relays emit their packets at the beginning of their assigned time slots. We assume their packet emission lasts for the whole time slot duration. Nodes that are not emitting in a time slot can receive packets in this time slot. Important to our model is that all relays that aren't emitting are listening. If packet reception is possible in the current time slot, the packet is stored in its incoming buffer.
%For instance, in the 2-relay network, if $S$ emits a packet in time slot 1, relays $A$ and $B$ try to receive this packet if are not transmitting in time slot 1.
In other words, in our graph $\mathcal{K_{|\V|}}$, if any node $i \in |\V|$ emits a packet in time slot $u$, all edges leaving $i$ in time slot $u$ ($(i,j,u)~\forall j\in \E$) carry the same packet to their next hops $j$.
As such, our model completely captures the broadcast property of the wireless medium.

We assume that relays have $|\T|$ incoming buffers and $|\T|$ outgoing buffers. All buffers are able to store the amount of packets transmitted in one time slot duration. In our examples, they can store one packet.
%A relay receiving a packet has to decide if it will discard it or in which slot of the next frame it will send it. This decision will be modeled as a forwarding probability in
We consider as well in our model that a relay can not differentiate packets: identical packets are indiscernible.

\subsubsection{Medium access control for broadcast transmissions}
We assume a very basic random channel access for all nodes sharing a time slot $u$: if a node $i$ is willing to transmit its packet in time slot $u$ in the next frame, it attempts it with probability $\tau_i^u$. The packet is disregarded with probability $1-\tau_i^u$. There is no acknowledgment procedure. If the receiver can not decode the packet, it is definitively lost. Contrary to more elaborated medium access procedures, emission decisions of nodes are independent.

An \emph{emission} is defined as the couple $(i,u) \in \V \times \T$ and represents the fact that node $i$ is emitting in a time slot $u$.
The \emph{emission rate} $\tau_i^u$ is defined as the probability node $i$ is emitting in time slot $u$.
Thus, a node $i$ transmitting at a rate of $\tau_i^u=0.5$ in time slot $u$ will decide with probability 0.5 to transmit its previously received packet in time slot $u$ in the upcoming frame. If this decision is successful, it will occupy time slot $u$ of the next frame for its whole duration.

Having this, a vector of emission rates for a node $i$ can be defined
$ \tau_i = \left[
	\begin{array}{ccc}
	\tau_i^1&
	 \dots&
	 \tau_i^{|\T|} \\
	 \end{array}
	 \right]$.
Let $\tau = \left[
 	\begin{array}{c c c}
		\tau_1^\dagger &
		\cdots &
		\tau_N^\dagger \\
 	\end{array}
\right]^{\dagger}
$ be the \emph{emission rate matrix}.
A particular instance of $\tau$ values \emph{is feasible} if and only if the following Properties 1 and 2 hold for each node:

{\sc Property 1:} {\it Flow conservation}.  The sum rate of all outgoing links is lower or equal to the sum rate of all incoming links, i.e.
\begin{equation}
  \overrightarrow{r_j} \leq \overleftarrow{r_j}  , ~~\forall j \in \V
\label{eq:ratesflowconservation}
\end{equation}
with
$
\overleftarrow{r_j} = \sum_{u \in \T} \sum_{(i,j) \in \Nout{j}{u}} \tau_i^u p_{ij}^u
$
the average rate at which all the packets are coming into node $j$ and
$
\overrightarrow{r_j} = \sum_{v \in \T} \tau_j^v
$ the rate at which packets are being transmitted by node $j$.

In the case of equality, we have a strict flow conservation. Otherwise, $| \overleftarrow{r_j} - \overrightarrow{r_j} |$ are dropped by node $j$ as a consequence of a forwarding decision described in the next item.

{\sc Property 2:} {\it Half duplex.} A node $j$ is able to receive a message on a time slot $u$ if it is not transmitting on that same time slot.
As a consequence, a $\tau$ is feasible if for each node of the network, the average number of time slots it spends transmitting and receiving sums up to a maximum value of one:
\begin{equation}
\overleftarrow{r_j^u} +  \tau_j^u \leq 1, ~~\forall (j,u) \in \V\times\T
\label{eq:halfduplex}
\end{equation}
\noindent where $\overleftarrow{r_j^u} = \sum_{(i,j) \in \Nout{j}{u}} \tau_i^u p_{ij}^u$ stands for the incoming cumulative rate in time slot $u$.

We define $\Gamma$ as the set of all feasible emission rate matrices.

\subsubsection{Forwarding and scheduling decisions}

Each node $j$ will decide, with the forwarding probability $x_{ij}^{uv}$, to transmit on time slot $v$ a packet coming from node $i$ in the time slot $u$ \emph{of the next frame}. Thus, we can define a $N|\T|$-by-$|\T|$ matrix giving all the forwarding probabilities relative to any node $j$ of the network. It is given by
$
X_j = \left[
 	\begin{array}{c c c}
		X_{1j} &
		\cdots &
		X_{Nj} \\
 	\end{array}
\right]^{\dagger}
$
where each matrix  $X_{ij}$ provides the scheduling probabilities of a flow of packets coming from node $i$ on its output times slots, depending on the time slot the packets are received on. We have
\[
 X_{ij} = \left[
 	\begin{array}{c c c}
		x_{ij}^{11} 	& \cdots & x_{ij}^{1|\T|}\\
		\vdots 	 	& 		&	\vdots	\\
		x_{ij}^{|\T|1} 	& \cdots & x_{ij}^{|\T||\T|}\\
 	\end{array}
\right].
\]
The matrix of forwarding probabilities is related to the matrix of emission rates $\tau$ and the matrix of channel probabilities $P$ with the following set of $|\M|$ equations
\begin{equation}
\sum_{(i,j) \in \Nout{j}{u}} \sum_{u \in \T}  \tau_i^u p_{ij}^u x_{ij}^{uv} =  \tau_j^v, ~~~\forall (j,v)\in \M
\label{eq:fwfprobaflowconservation}
\end{equation}
\noindent where $\tau_i^u p_{ij}^u$ is the probability that a packet sent by $i$ on time slot $u$ arrives in $j$.
These equations are derived from the flow conservation property of \eqref{eq:ratesflowconservation}. They strictly constrain the choices of forwarding probabilities.

The forwarding probabilities represent the decisions of the nodes to either $(i)$ retransmit all the packets received or $(ii)$ reduce the output rate by dropping or re-encoding them together.
From now on, we will refer to the set of all forwarding probabilities of the complete network using a matrix $X=\left[ X_1 \dots X_N \right],~ X \in \X$ of size $N.|\T|$-by-$N.|\T|$ where $\X$ is the set of all possible matrix instances.

\section{MO optimization problem}\label{sec:problemstatement}

\subsection{Elementary criteria definition}\label{sec:steadystate}

This section defines for one source-destination flow optimization objectives related to reliability, capacity, end-to-end delay and energy consumption based on the aforementioned network and protocol model.

{\sc Capacity objective $f_C$:} It is defined as the average number of packets received by the destination per packet sent by $S$. If $\mathcal{P}$ is the set of all possible paths on $\mathcal{K_{|\V|}}$ between $S$ and $D$, it is derived by summing the transmission success probability of a packet on each path. Formally:
\begin{equation}
f_C = \sum_{\paths \in \mathcal{P}} P(\paths)
\label{eq:capacite}
\end{equation}
where $P(\paths)$ is the transmission success probability of a packet on path $\paths \in \mathcal{P}$.

{\sc Reliability objective $f_R$:} It is defined as the probability of a packet to arrive at the destination. It is equivalent to the success rate of a packet sent by the source. It differs from the capacity criterion because redundant packet copies that successfully arrive at the destination are not accounted for. More specifically, it is the probability that at least one copy arrives at $D$. Formally:
\begin{equation}
f_R = 1-\prod_{\paths \in \mathcal{P}} 1-P(\paths)
\end{equation}

{\sc Delay objective $f_D$:}  It is defined as the average delay a packet sent by the source needs to reach the destination, expressed in number of hops.
Assuming that one hop introduces a delay of 1 unit, a $h$-hop transmission introduces a delay of $h$ units.
Having $H(\paths)$ the length in hops of path $\paths$, the average end to end delay is computed by:
\begin{equation}
\begin{split}
f_D = \frac{\sum_{\paths \in \mathcal{P}} H(\paths)\cdot P(\paths)}{\sum_{\paths \in \mathcal{P}} P(\paths)} = \frac{\sum_{\paths \in \mathcal{P}} H(\paths)\cdot P(\paths)}{f_C}
\end{split}\label{eq:delay}
\end{equation}
where the numerator provides the total delay of all paths and the denominator the number of copies received, in average.

{\sc Energy objective $f_E$:}
We consider as a first approximation that the main energy consumption factor is due to the emission of a packet. Thus, the energy criterion $f_E$ is defined as the average number of emissions performed by all nodes (source and relays) per packet sent by the source. This simple energy model will be improved in future works to account for idling, listening and receiving energy expenditure.

\subsection{Capacity and reliability achieving criteria}
We define as well two other types of criteria, naming \emph{reliability achieving} and \emph{capacity achieving} criteria.
These objectives directly derive from the elementary objectives introduced earlier.

{\sc Reliability achieving delay $f^r_D$ and energy $f^r_E$:}
Reliability-achieving delay and energy criteria are defined as follows:
\begin{eqnarray}
f^r_D=f_D/f_R\\
f^r_E=f_E/f_R
\end{eqnarray}
They represent the delay and energy needed to reach a perfectly reliable transmission.
For instance, if  $f_C=0.5$, $f^r_D=2 f_D$ and $f^r_E = 2 f_E$, meaning that 2 times more packets have to be sent in average to reach perfect reliability at the cost of double delay and energy.

{\sc Capacity achieving delay $f^c_D$ and energy $f^c_E$:}
Capacity-achieving delay and energy criteria are defined as follows:
\begin{eqnarray}
f^c_D=f_D/\min(f_C,1)\\
f^c_E=f_E/\min(f_C,1)
\end{eqnarray}
Here, capacity achieving criteria  $f^c_D$ and $f^c_E$ are obtained by dividing the value of $f_D$ and $f_E$ by min$(f_C,1)$ respectively.

Capacity and reliability criteria are equal if the packet travels on a unique path between $S$ and $D$.
When more than one path connect $S$ and $D$, these criteria are not equal anymore because several copies may reach $D$.
More generally, $f_C$ upper bounds reliability: $f_C \geq f_R$.

If $f_R=1$, it implies that $f_C\geq 1$ but the converse is not true.
For instance, if two paths with non-null transmission probabilities exist between $S$ and $D$ and if $f_C=1$, either zero, one or two copies of the original packet can be received at $D$, with a temporal average of 1 packet per frame.
The cases where zero or two copies are received are not interesting of course. But we show in Section \ref{sec:coding} that this capacity criterion can be reached if relays perform network coding, introducing diversity into the packets they are relaying.

\subsection{Pareto-optimal bound and solution set}

The first goal of this paper is to derive the \emph{Pareto-optimal performance bounds} and the set of corresponding \emph{Pareto-optimal networking solutions} of the considered network topologies with respect to given performance objectives.

\subsubsection{Pareto-optimality}
Formally, a Pareto-optimal solution set is composed of all the \emph{non-dominated solutions} of the MO problem with respect to the performance metrics considered.
The definition of dominance is:

{\sc Definition 1:} A solution $x$ dominates a solution $y$ for a $n-$objective MO problem if $x$ is at least as good as $y$ for all the objectives and $x$ is strictly better than $y$ for at least one objective.

{\sc Definition 2:} A solution $x\in \solSpace$ is Pareto-optimal if there is no other solution $y\in \solSpace$ that dominates $x$.

Thus, the \emph{set of Pareto-optimal solutions} is as follows:
\begin{equation}
\solSpace_{opt} = \{x \in \solSpace~: ~\forall y \in \solSpace_{opt},  y~\mathrm{does~not~dominate}~x\}
\label{eq:MOproblem}
\end{equation}
and the corresponding \emph{Pareto-optimal performance bound} is:
\begin{equation}
\bound_{opt} = \{\left(f_1(x), f_2(x), \dots,  f_n(x)\right)~\forall x \in \solSpace_{opt}\}
\label{eq:MOproblem}
\end{equation}
with $f_1$, $f_2$, $\dots f_n$ the $n$ arbitrary objective functions.

\subsubsection{Solution set}
Based on our network model, several parameters can be treated as optimization variables: the location of the $N$ relays, the number of relays $N$, the transmission power of each relay, the number of time slots $|\T|$ and the forwarding probabilities represented by matrix $X \in \X$.
In this paper, considered variables are the location of the $N$ relays and their respective forwarding probabilities.
Location $l$ of $N$ relays can be chosen in a convex set $l \in \mathcal{C}^N$.
For one network realization, the forwarding probabilities are chosen in the set $\X$.
Thus the complete search space is $\mathcal{S} = \mathcal{C}^N\times\X$.

\subsection{MO Optimization problem}
Even if the forwarding probabilities are the main optimization variables, we have to derive the emission rate matrix to exactly compute the interference level in the network.
As shown in \cite{Jaffres-Runser2010-WiOpt}, the derivation of the emission rate matrix $\tau \in \Gamma$ knowing the forwarding probabilities $X \in \X$ is intractable. In a nutshell, to compute the emission rates of a node $j$ knowing $X$,  the incoming transmission probabilities $p_{ij}^u$ have to be known as well. Yet, in order to compute the $p_{ij}^u$ values, the emission rates of the nodes $i \in \V, i \neq j$ are needed, creating a circular dependency between the emission rates.

Thus, we have proposed a reverse approach in \cite{Jaffres-Runser2010-WiOpt} where the main optimization variable is switched to the set of feasible emission rate matrices $\Gamma$. From any feasible value $\tau \in \Gamma$, it is straightforward to derive the channel probabilities since the activity of all nodes on each time slot is known.

Only instances of $\tau$ that meet the constraints relative to Property 1 and 2 are further considered as valid.
Now that we have a valid $\tau$, we can derive all the forwarding matrices $X \in \X$ that verify the constraints of equation \eqref{eq:fwfprobaflowconservation}.
There are $|\M|$ constraints, each one constraining the choice of the $x_{ij}^{uv}$ for all nodes and time slots of the network with respect to $\tau$.
Let $\X^{\tau}$ be the subset of $\X$ that verifies \eqref{eq:fwfprobaflowconservation} with respect to the emission rate matrix $\tau$.
Each solution $X \in \X^{\tau}$ can be evaluated according to $f_C, f_D, f_R$ or $f_E$.

In the rest of the paper, the following multiobjective optimization problem that concurrently maximizes capacity and minimizes end-to-end delay and overall energy consumption is solved:
\begin{eqnarray}
\label{eq:MOproblem}
  & \left[ \max f_C(x), \min f_D(x), \min f_E(x)\right]^T  \label{eq:MOproblem} \\ \nonumber
s.t. & \\ \nonumber
x = & (l, \tau, X) \in  \mathcal{C}^N\times\Gamma \times \X^{\tau}
\end{eqnarray}
where $l$ is the location vector of $N$ relays, $\tau$ a feasible emission rate matrix and $X$ a feasible forwarding matrix for $\tau$.

Additional constraints can be included in this MO problem depending on the type of analysis needed or to reduce the size of the search space. For instance, nodes can be assigned time slots beforehand. In this case, if channel $u$ is not assigned to node $i$, the $\tau_i^u$ variable becomes a constant equal to zero. The $\tau_i^u=0$ constraint would then be added to the MO problem.

\subsubsection{Multiobjective performance bounds}
Three different types of bounds are investigated in this work, derived from the solution of the MO optimization problem defined in Eq.~\eqref{eq:MOproblem}.

The first bound, referred to as $\bound_{opt}$, is directly obtained by solving Eq.~\eqref{eq:MOproblem}.
It is defined as:
\[
\bound_{opt} = \{\left(f_C(x), f_D(x), f_E(x)\right)~\forall x \in \solSpace_{opt}\}
\]
where $\solSpace_{opt}$ is the corresponding Pareto-optimal solution set.

From $\bound_{opt}$, the two following bounds are derived:
\begin{itemize}
\item The capacity-achieving \emph{upper} bound: \[\bound^c = \{\left(f^c_D(x), f^c_E(x)\right)~\forall x \in \solSpace_{opt}\}\]
\item The reliability-achieving \emph{lower} bound: \[\bound^r = \{\left(f^r_D(x), f^r_E(x)\right)~\forall x \in \solSpace_{opt}\}\]
\end{itemize}
Both bounds are calculated by applying capacity and reliability achieving criteria to the solutions of $\solSpace_{opt}$.

The MO problem of Eq.~\eqref{eq:MOproblem} optimizes the capacity criterion and not the reliability criterion.
As presented earlier, the capacity criterion upper bounds the reliability criterion.
The reliability-achieving bound $\bound^r$ is a feasible delay-energy performance bound obtained as relays forward packets according to the forwarding probabilities of the solutions of $\solSpace$.
On the contrary, the capacity-achieving bound $\bound^c$ is an upper bound because it accounts for multiple copies.
We show in Section \ref{sec:coding} that it is possible to reach the capacity-achieving bound using network coding, and that this bound is tight for the 1 and 2-relay networks.

These bounds are not Pareto-optimal with respect to $(f^c_D, f^c_E)$ or $(f^r_D, f^r_E)$.
From $\bound^c$ (resp. $\bound^r$) the set of non-dominated solutions with respect to  $f^c_D$ and $f^c_E$ (reps. $f^r_D$ and $f^r_E$) is selected. These sets are referred to as:
\begin{itemize}
\item $\bound^c_{opt}$, the Pareto capacity-achieving upper bound,
\item $\bound^r_{opt}$, the Pareto reliability-achieving lower bound.
\end{itemize}

\section{Pareto bounds for 1-relay and 2-relay topologies}\label{sec:1relay2relay}

\subsection{Study cases}
For the two topologies presented in Fig.~\ref{fig:topologies}, five different study cases are considered and summarized in Table \ref{tab:studyCases}.
In all study cases, the source only emits packets on the first time slot with rate one: $\tau_S^1 = 1$ and $\forall u \neq 1, \tau_S^u = 0$.

Each study case defines which topology is assumed, how many time slots constitute a frame and which nodes are allowed to transmit in each time slot.
The time slots assignments of study cases 1, 3 and 4 ensure no interference exists, while other study cases exhibit time slots with possible interference.
For the 2-relay topology, transmissions between relays $A$ and $B$ are possible or not. If they are possible, packets may loop infinitely between $A$ and $B$.
All study cases are defined by introducing additional constraints into the MO optimization problem defined in \eqref{eq:MOproblem}.
For each study case, section \ref{subsec:MOproblem} defines the exact MO optimization problems solved.

\begin{table}
\begin{tabular}{c c  c c c c c}
Study 	& 			&  		& \multicolumn{3}{c}{Nodes transmitting on slot} & Loop between \\
case 	& Topology 	& $|\T|$ 	& 1 	&  2		& 3 	& $A$ and $B$\\ \hline\hline
1 & 1-Relay		& 2 								& $S$		& $R$ 	& - & - \\ \hline
2 & 1-Relay		& 1 								& $S,R$ & - & - & - \\ \hline
3 & 2-Relay		& 3 								& $S$		& $A$ 	& $B$ & No\\ \hline
4 & 2-Relay		& 3 								& $S$		& $A$ 	& $B$ & Yes\\ \hline
5 & 2-Relay		& 2 								& $S$		& $A,B$ 	& - & No\\ \hline
%6 & 2-Relay		& 2 								& $S$		& $A,B$ 	& - & Yes\\ \hline
\end{tabular}
\caption{Study cases}\label{tab:studyCases}
\end{table}

\subsection{Criteria for 1-relay and 2-relay topologies}\label{subsec:MOproblem}
For both topologies, the general multiobjective problem of \eqref{eq:MOproblem} is considered.
Expressions for the optimization criteria of each study case are detailed in this subsection.

\subsubsection{1-relay topology}
For a 1-relay topology using $|\T|$ time slots, optimization objectives are defined as:
\begin{eqnarray}\label{eq:fc1}
f_C  = & \tau_S^1p_{SD}^1 + \sum_{u=1}^{|\T|} \tau_S^1p_{SR}^1x_{SR}^{1u}p_{RD}^{u}\\
f_R  = & 1- (1-\tau_S^1p_{SD}^1)\prod_{u=1}^{|\T|} \left( 1- \tau_S^1p_{SR}^1x_{SR}^{1u}p_{RD}^{u}\right)\\
%f_D  = & 1/f_C \cdot \sum_{u=1}^{|\T|}\tau_S^1p_{SR}^1x_{SR}^{1u}p_{RD}^{u}\\
f_D  = & ( \sum_{u=1}^{|\T|}2 \cdot \tau_S^1p_{SR}^1x_{SR}^{1u}p_{RD}^{u} + \tau_S^1.p_{SD}^1)/f_C\\
f_E  = & \tau_S^1 +  \sum_{u=1}^{|\T|} \tau_S^1p_{SR}^1x_{SR}^{1u} \label{eq:fc4}
\end{eqnarray}

Briefly, $f_C$ adds the probability for a packet in arrive to $D$ directly (i.e. $\tau_S^1p_{SD}^1$) and the probability for the same packet to arrive through the relay node for each available time slot. The reliability criterion equals the probability for at least one packet to arrive through any available path and time slot.
The delay criterion sums the delays of all packets arriving through all possible paths and averages it with the value of $f_C$.
Finally, the energy criterion sums the number of emissions of the source and the relay, knowing the probability that the relay will receive and forward packets.

Study cases 1 and 2 are covered by the MO problem of \eqref{eq:MOproblem} and objectives are given in Eq.~\eqref{eq:fc1} to \eqref{eq:fc4}.
For {\bf study case 1}, $|\T|=2$ times slots, $S$ emits in time slot 1 and relay $R$ in time slot 2. Additional constraint to MO problem \eqref{eq:MOproblem} is $\tau_R^1=0$.
Thus, following \eqref{eq:fwfprobaflowconservation}, $x_{SR}^{11}=0$ and $x_{SR}^{12}= \tau_R^2 / (\tau_S^1.p_{SR}^1)$. As such, the only variables in this problem are the location of the relay $l_R \in \mathcal{C}$ and its forwarding probability $x_{SR}^{12}\in [0,1]$.

For {\bf study case 2}, since there is only one time slot, a single variable $x_{SR}^{11}$ is defined.
It is directly related to $\tau_R^1$ following \eqref{eq:fwfprobaflowconservation}: $x_{SR}^{11}=\tau_R^1/(\tau_S^ 1p_{SR}^1)$.  As such, the only variables in this problem are the location of the relay $l_R \in \mathcal{C}$ and its forwarding probability $x_{SR}^{11}\in [0,1]$.

\subsubsection{2-relay topology}
Study cases 3 to 5 are covered by the MO problem of \eqref{eq:MOproblem}. Objectives and additional constraints are defined hereafter.

For the interference free study cases 3 and 4, we have $|\T|=3$ time slots. $S$ is still transmitting in time slot 1 while relays $A$ and $B$ are transmitting in time slot 2 and 3, respectively. Following from the slot allocation, only $\tau_A^2$ and $\tau_B^3$ are defined, other relay emission rates are set to 0. From Eq.~\eqref{eq:fwfprobaflowconservation}, we deduce that only $x_{SA}^{12}$, $x_{SB}^{13}$, $x_{BA}^{32}$ and $x_{AB}^{23}$ are non zero variables. Moreover, an $X \in \X$ matrix is feasible if the 2 constraints on the forwarding probabilities originating from \eqref{eq:fwfprobaflowconservation} are met:
\begin{equation*}
\begin{array}{c}
    \tau_S^1.p_{SA}^1.x_{SA}^{12} + \tau_B^3.p_{BA}^3.x_{BA}^{32} = \tau_A^2 \\
    \tau_S^1.p_{SB}^1.x_{SB}^{13} + \tau_A^2.p_{AB}^2.x_{AB}^{23} = \tau_B^3 \\
\end{array}
\end{equation*}

Introducing the notation  $\Qe{i}{j}{u}{v}=p_{ij}^ux_{ij}^{uv}$, the optimization objectives are:
\begin{equation}
\begin{split}
f_C  =  \tau^1_S.p_{SD}^{1} + \frac{\tau_S^1}{1-Q_{AB}^{23}Q_{BA}^{32}}.(E + F)
%\tau_S^1 [p_{SD}^1 + \frac{1}{} \\ \left[Q_{SA}^{12}(p_{AD}^2+Q_{AB}^{23}p_{BD}^3)+Q_{SB}^{13}(p_{BD}^3+Q_{BA}^{32}p_{AD}^2) \right]]
\end{split}\label{eq:fc}
\end{equation}
with $E = (\Qe{S}{A}{1}{2} + \Qe{S}{B}{1}{3}.\Qe{B}{A}{3}{2}).p_{AD}^{2}$ and $F = (\Qe{S}{B}{1}{3} + \Qe{S}{A}{1}{2}.\Qe{A}{B}{2}{3}).p_{BD}^{3}$.
\begin{equation}
\begin{split}
%f_D  = \frac{1}{f_C} \cdot \frac{\tau_S^1}{(1-Q_{AB}^{23}Q_{BA}^{32})^2}(A+B)
f_D  = \frac{1}{f_C} \cdot [\frac{\tau_S^1}{(1-Q_{AB}^{23}Q_{BA}^{32})^2}(A+B) + \tau_S^1.p_{SD}^1]
\end{split}\label{eq:fd}
\end{equation}

%\noindent with $A = p_{AD}^2[\Qe{S}{A}{1}{2}(1+\Qe{A}{B}{2}{3}.\Qe{B}{A}{3}{2})+2\Qe{S}{B}{1}{3}.\Qe{B}{A}{3}{2}]$ and $B =p_{BD}^{3}[\Qe{S}{B}{1}{3}(1+\Qe{A}{B}{2}{3}.\Qe{B}{A}{3}{2})+2\Qe{S}{A}{1}{2}.\Qe{A}{B}{2}{3}]$.

\noindent with $A = p_{AD}^2[\Qe{S}{B}{1}{3}.\Qe{B}{A}{3}{2}(3-\Qe{A}{B}{2}{3}.\Qe{B}{A}{3}{2})+2\Qe{S}{A}{1}{2}]$ and $B =p_{BD}^{3}[\Qe{S}{A}{1}{2}.\Qe{A}{B}{2}{3}(3-\Qe{A}{B}{2}{3}.\Qe{B}{A}{3}{2})+2\Qe{S}{B}{1}{3}]$.

\begin{equation}
\begin{split}
f_E =\tau_S^1 + \frac{\tau_S^1}{1-Q_{AB}^{23}Q_{BA}^{32}}(\Qe{S}{A}{1}{2} +  \Qe{S}{B}{1}{3}.\Qe{B}{A}{3}{2} \\+ \Qe{S}{A}{1}{2}.\Qe{A}{B}{2}{3} + \Qe{S}{B}{1}{3})
\end{split} \label{eq:fe}
\end{equation}

Detailed derivation of these criteria are presented in the Appendix. These equations originate from infinite summations over all possible path lengths. Indeed, due to the loop, packets may travel up to an infinite number of hops in the network. If $Q_{AB}^{23}Q_{BA}^{32} <1$, the geometric serie of ratio $Q_{AB}^{23}Q_{BA}^{32}$ with first term 1 converges and finite values for $f_C$, $f_D$ and $f_E$ can be derived. Expressing $f_R$ as a function of this infinite sum is not possible and this criterion is evaluated through simulations when a loop exists between $A$ and $B$.

$Q_{AB}^{23}Q_{BA}^{32}$ may be equal to one if perfect links between $A$ and $B$ exist and $x_{AB}^{23}=x_{BA}^{32}=1$. As such, we add the following constraint to the MO problem of \eqref{eq:MOproblem} for study case 4: %cases 4 and 6:
\begin{equation*}
\forall X \in \X ~\mathrm{if}~(p_{AB}^2=1 \wedge p_{BA}^3=1) \left\{
\begin{array}{c}
x_{AB}^{23}<1-\Delta \\
x_{BA}^{32}<1-\Delta \\
\end{array}
\right.
\end{equation*}
with an empirically chosen value of $\Delta = 0.05$.

For study cases 3 and 5, additional constraints that avoid relay $A$ to forward packets from $B$ and vice versa are defined: $x_{AB}^{23}=x_{AB}^{32}=0$ for case 3 and  $x_{AB}^{22}=x_{AB}^{22}=0$ for case 5. Only one-hop and two-hop transmissions are possible.
$f_C$, $f_D$ and $f_E$ can be deduced from \eqref{eq:fc}, \eqref{eq:fd}  and \eqref{eq:fe}, respectively. In both study cases 3 and 5, it is possible as well to derive a closed form expression for $f_R$:
\begin{equation}\label{eq:fr}
\begin{split}
f_R  = 1- (1-\tau_S^1p_{SD}^1)( 1- \tau_S^1p_{SA}^1x_{SA}^{12}p_{AD}^{2})\\
\times( 1- \tau_S^1p_{SB}^1x_{SB}^{13}p_{BD}^{3})\\
\end{split}
\end{equation}

Study cases 5 %and 6 differ
differs from 3 and 4 respectively by its time slot assignment. Indeed, $A$ and $B$ emit on the same slot 2. As such, aforementioned criteria are straightforward to adapt to this other time slot assignment.

\subsection{Implementation}\label{subsec:Implementation}
\begin{figure}[h]
\begin{center}
\scalebox{0.4}{
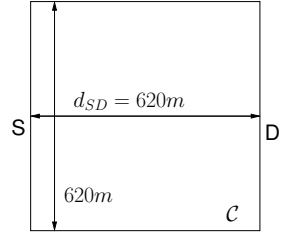
}
\caption{Relay location search space $\mathcal{C}$ }
\label{fig:convex}
\end{center}
\end{figure}

For most of the study cases, the distance between $S$ and $D$ is set to $d_{SD}=620 m$ such as having a direct transmission probability $p_{SD}^1$ without interference near 0 (assuming a transmission power $\power=0.15$mW and a pathloss exponent of 3).
The set of Pareto optimal locations of relays is searched in a continuous square surface area $\mathcal{C}$ of size $d_{SD}\times d_{SD}$ meters as shown in Figure~\ref{fig:convex}.

When study cases where different $|\T|$ values are compared on the same figure, they are scaled to be comparable.

\subsubsection{MO optimization}
Theoretical Pareto-optimal solutions and bounds are obtained using the state of the art non-dominated sorting genetic algorithm (NSGA-2) \cite{Deb2002}.
For NSGA-2, a population size of 300 solutions is used and a maximum number of 1000 generations. The crossover probability is set to 0.9.

For each study case, upper Pareto bounds $\bound_{opt}$ between capacity, delay and energy objectives are computed analytically by solving the corresponding MO optimization problem.
Then, the capacity-achieving upper bound $\bound^c$ is calculated analytically. The reliability-achieving lower bound $\bound^r$ is calculated analytically for the no-loop study cases and empirically using simulations otherwise.

\subsubsection{Simulations settings}
To assess analytical results, both upper and lower bounds are compared with simulations obtained with the event-driven network simulator WSNet\footnote{http://wsnet.gforge.inria.fr/} \cite{WSNet}.

Each upper Pareto bound solution is simulated with WSNet. For each solution, the location of the relays and their forwarding probabilities are known. These forwarding probabilities are used in simulation to decide whether to forward  a packet upon its reception. If the decision of packet emission is not successful, the packet is disregarded.
In our simulations, a perfect TDMA is implemented following the specifications of the study case of interest. $S$ sends a packet every first time slot of every frame. Experiment is run for 10000 frames.

Simulated values $\tilde f_C$, $\tilde f_D$, $\tilde f_E$, $\tilde f^c_D$ and $\tilde f^c_E$ are calculated from simulations as follows.
$\tilde f_C$ is measured by the total number of packets $N_{rx}$ received at $D$ (including copies) divided by the number of packets transmitted by $S$.
$\tilde f_D$ is measured using the statistical distribution of the delays of the packets arrived over each possible distance measured in hops:  $P(h)=n(h)/N_{rx}$, where  $n(h)$ is the number of packets arrived in $h$ hops at $D$. $\tilde f_D$ is then calculated with $\tilde f_D=\sum_{h=1}^{h_{max}} h\cdot P(h)$, with $h_{max}$ the maximum number of hops of all packets collected at $D$.
$\tilde f_E$ is the sum of the number of packets transmitted by the source and the relays divided by the number of packets sent by $S$.

Besides, for each study case we compute empirically the reliability objective $f_R$. It measures the proportion of different packets arriving at $D$. It is a regular success rate (copies are disregarded).
From $\tilde f_R$, empirical \emph{reliability-achieving} delay $\tilde f^r_D = \tilde f_D/\tilde f_R$ and energy $\tilde f^r_E=\tilde f_E/\tilde f_R$ objectives are defined.

\section{Results for considered topologies}\label{sec:results}

This section presents the bounds $\bound_{opt}$, $\bound^c$ and $\bound^r$ computed analytically and by simulations for the study cases introduced earlier. We extract as well the $\bound^c_{opt}$ and $\bound^r_{opt}$ bounds representing the set of non-dominated solutions of $\bound^c$ and $\bound^r$ with respect to capacity-achieving and reliability-achieving criteria, respectively.

A first important conclusion is that in all figures, the analytical bounds and their simulated counterparts perfectly match, assessing our network model and criteria definitions. Table \ref{tab:RMSEnoNC} gives the root mean square error (RMSE) between $\bound_{opt}$ and $\tilde \bound_{opt}$ for all study cases. $RMSE= \frac{1}{N_{opt}}\sqrt{\sum_{i=1}^{N}\frac{(f(i)-\tilde f(i))^2}{f(i)^2}}$
where $N_{opt}$ is the total number of Pareto-optimal solutions in $\bound_{opt}$. Values are really small, showing a quasi-perfect match between the model and simulations.
\begin{table}
\caption{The RMSE and generational distance for study cases}
\centering
\label{tab:RMSEnoNC}
\begin{tabular}{|l|l|l|l|} \hline
Study Cases & $f_C$   & $f_D$ & $f_E$ \\ \hline
Study case 1($p_{SD}^1\simeq 0.5$) & 4.6e-05 & 1.5e-05 & 1.5e-05\\
Study case 1($p_{SD}^1\simeq 0$) & 1.7e-03 & 0 & 1.8e-05\\ \hline
Study case 2($p_{SD}^1\simeq 0.5$) & 4.3e-05 & 1.8e-05 & 1.1e-05\\ \hline
%Study case 2($p_{SD}^1\simeq 0$) & 8.4e-06 & 6.9e-06 & 5.2e-05\\ \hline
Study case 3 & 1.6e-04 & 0 & 2.1e-05\\  \hline
Study case 4 & 5.2e-04 & 1.1e-03 & 2.2e-04\\  \hline
Study case 5 & 2.6e-04 & 0 & 1.8e-05\\ \hline
%Study case 6 & 1.3e-03 & 2.5e-04 & 2.5e-02\\ \hline
\end{tabular}
\end{table}

\subsection{1-relay Pareto bounds and sets}
Study cases 1 and 2 are investigated for two configurations: $p_{SD}^1\simeq 0$ ($d_{SD}=620$m) and $p_{SD}^1\simeq 0.5$ ($d_{SD}=310$m).
Bounds $\bound_{opt}$, $\bound^c$ and $\bound^r$ are given in Figures~\ref{fig:sc1d620} and \ref{fig:sc1d310} for study cases 1 and in Figure~\ref{fig:sc2d310}  for study case 2.

Another important conclusion is that when the source is connected to $D$ with a single path, there is a perfect match between $\bound^c$ and $\bound^r$ as expected. This is true for study cases 1 and 2 where $p_{SD}^1\simeq 0$ since only one path from $S$ to $D$ exists. When multiple copies arrive at the destination, $f_C$ and $f_R$ are different by definition, creating different bounds $\bound^c_{opt}$ and $\bound^r$.

Lastly, a clear compromise is visible: decreased energy is obtained at the price of an increase in delay. Solutions that consume less energy have the relay forward packets with a lower probability, creating less reliable communication. Less reliable solutions introduce an extended delay to achieve perfect capacity as shown by the increase in capacity or reliability-achieving delay.
A detailed explanation of the results presented for study cases 1 and 2 is given hereafter.

\subsubsection{\sc Study Case 1}  No interference exits in this scenario.

\begin{figure}[h]
\begin{center}
\scalebox{0.75}{
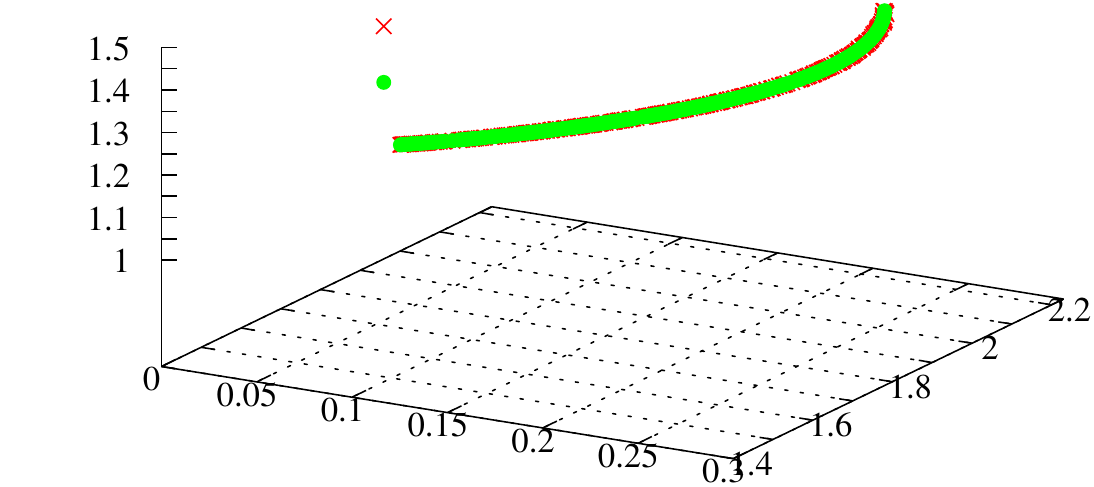
}
\vskip7pt
\scalebox{0.6}{
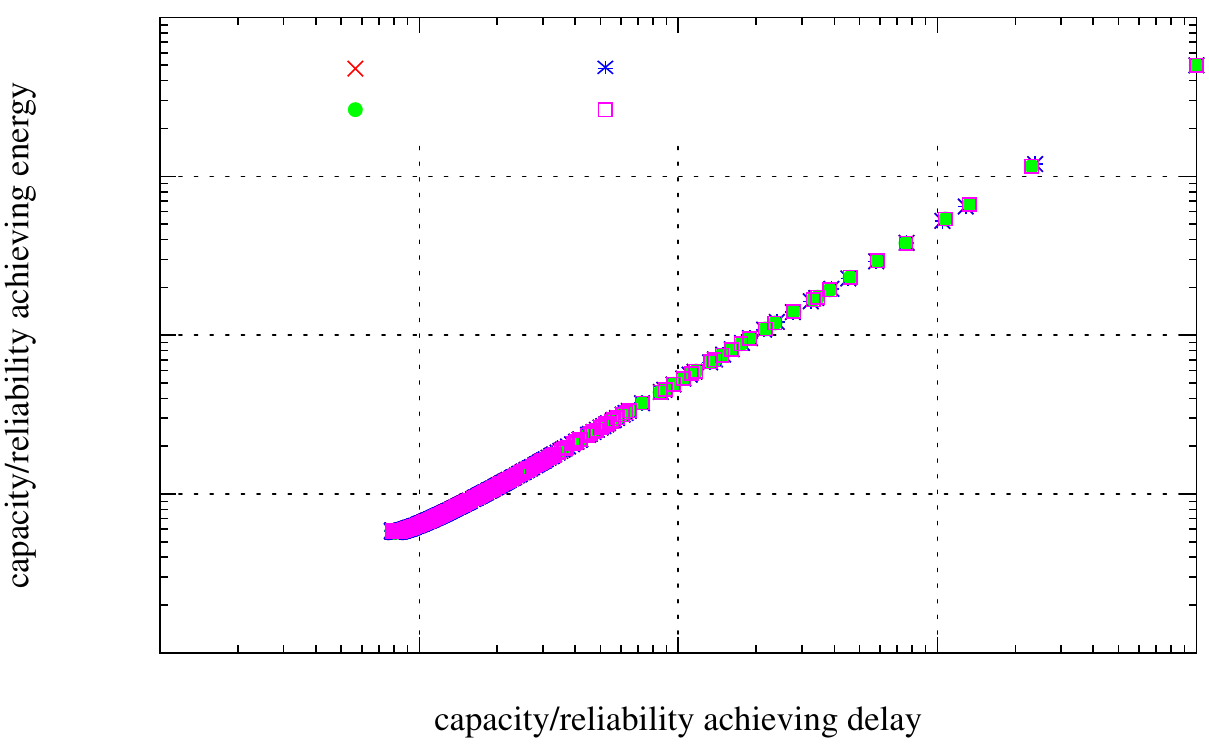
}
\vskip7pt
\scalebox{0.6}{
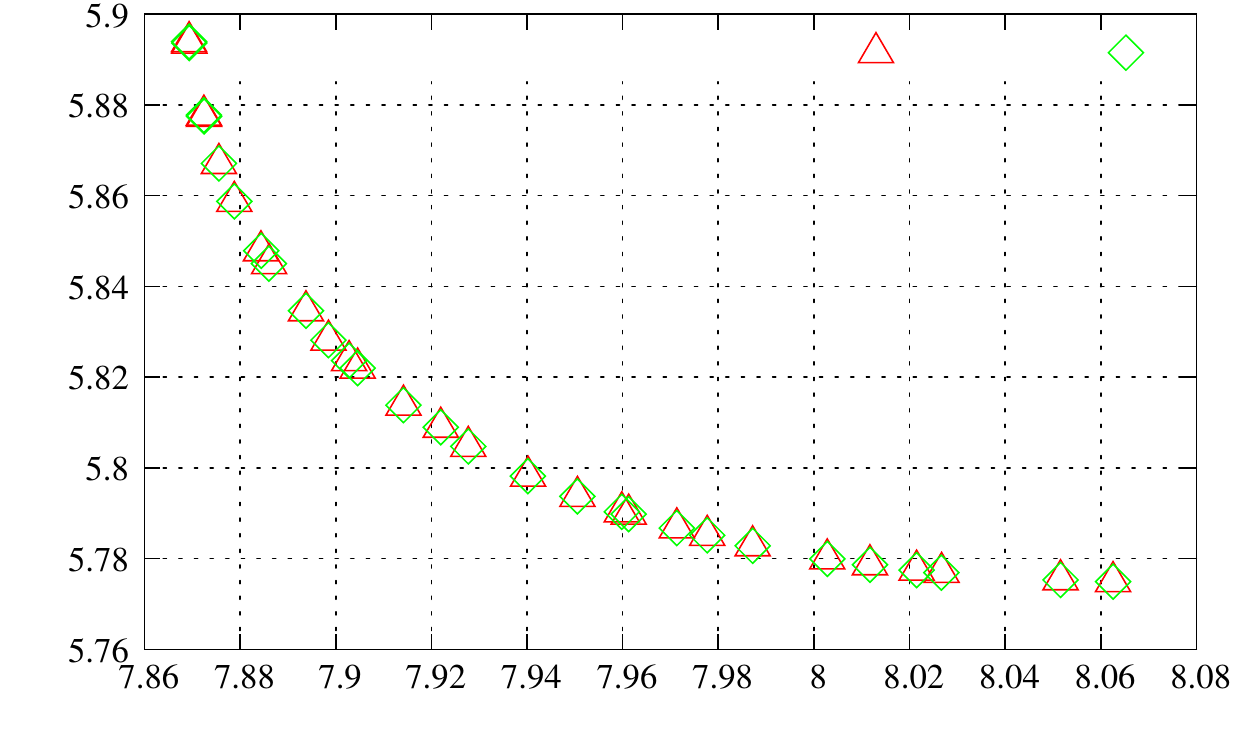
}
\caption{{\sc Study Case 1} with $p_{SD}^1\simeq 0$: Pareto-optimal bound $\bound_{opt}$ (top). Capacity and reliability-achieving bounds $\bound^c$ and $\bound^r$ (middle) and their corresponding Pareto-optimal bounds  $\bound^c_{opt}$ and $\bound^r_{opt}$ (bottom).}
\label{fig:sc1d620}
\end{center}
\end{figure}

\begin{figure}[t]
\begin{center}
\scalebox{0.7}{
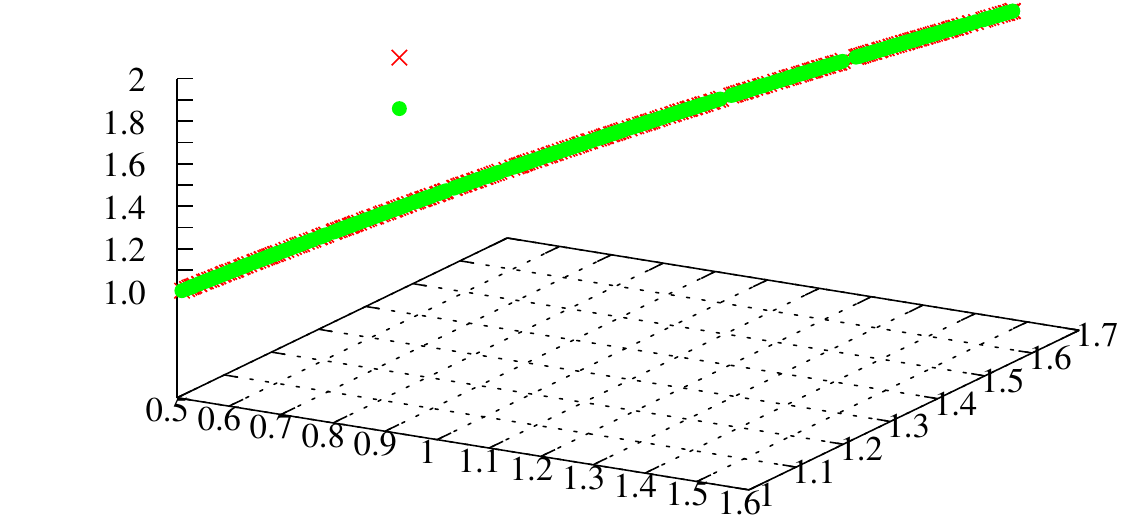
}
\vskip10pt
\scalebox{0.6}{
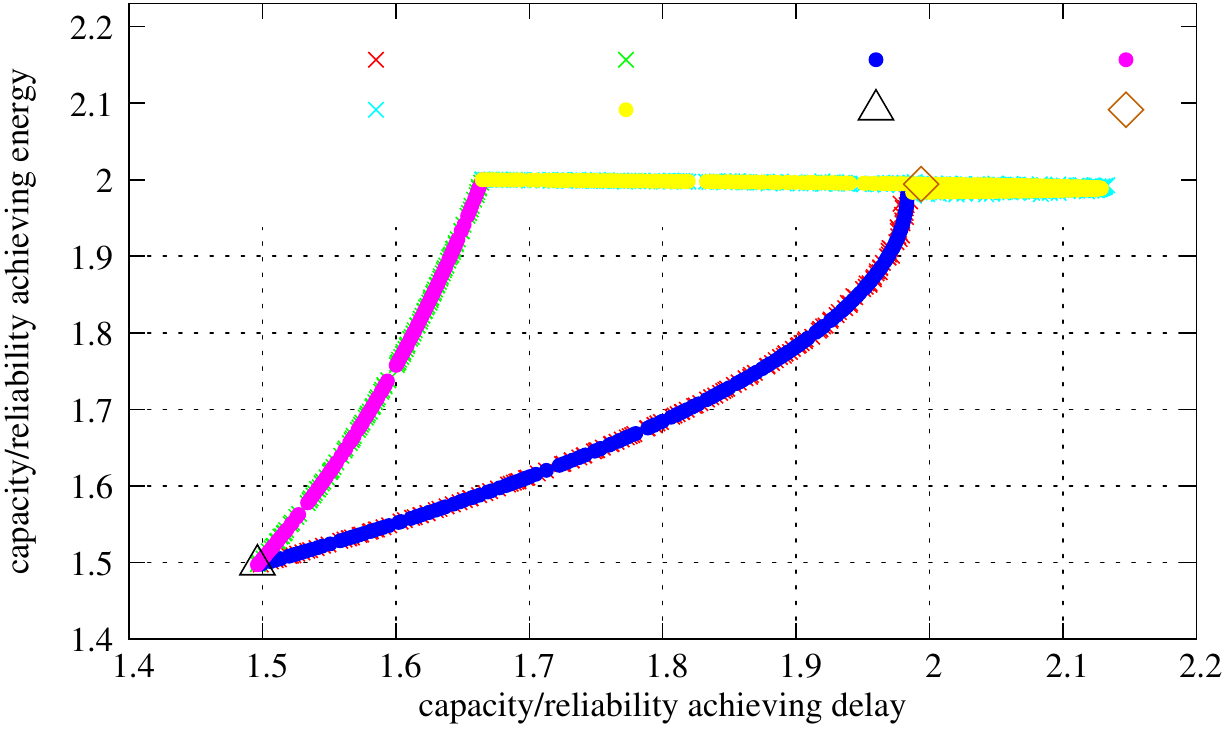
}
\caption{{\sc Study Case 1} with $p_{SD}^1\simeq 0.5$:  Pareto-optimal bound $\bound_{opt}$ (top). Capacity and reliability-achieving bounds (bottom). }
\label{fig:sc1d310}
\end{center}
\end{figure}

For $p_{SD}^1\simeq 0$, all bounds are given in Fig.~\ref{fig:sc1d620}. $\bound_{opt}$ is presented in Fig.~\ref{fig:sc1d620}-(top). For this bound, the solution with lowest capacity ($f_C \simeq 0$), lowest delay ($f_D=2$) and lowest energy ($f_E = 1$) is experienced when the relay node is \emph{not} contributing to the transmission. This is the case if the relay's forwarding probability $x_{SR}^{12}$ is zero or if the relay is not covered by $S$ ($p_{SR}^1\simeq 0$). Energy is minimized in this case since the relay never re-transmits packets. A delay of 2 is experienced because since $p_{SD}^1=0$ and the path probability on the $S-R-D$ path ($\tau_S^1p_{SR}^1x_{SR}^{12}p_{RD}^{2}$) is very small but not null, the few packets arriving have a delay of two. More generally, we can say that for any non zero value of $\tau_S^1p_{SR}^1x_{SR}^{12}p_{RD}^{2}$, which is equal to $f_C$, the delay is always equal to 2.

For the Pareto-optimal solution with highest capacity ($f_C =0.25$), highest delay ($f_D =2$) and highest energy ($f_E = 1.5$), the relay is located right in the middle of the $[S,D]$ segment and forwarding all received packet ($x_{SR}^{12}=1$). In this case, link probabilities between $S$ and $R$ and between $R$ and $D$ are maximized ($p_{SR}^1\simeq 0.5$ and $p_{RD}^2\simeq 0.5$), providing maximum energy consumption and capacity. Delay is still equal to two hops because each packet arrives on the $S-R-D$ path.
For the solutions that lie in between highest and lowest Pareto solutions, as $f_C$ is getting smaller, the relay is getting closer to $D$. In this case, the $S-R$ link has a weaker link while the $R-D$ link has a stronger link. $f_E$ is getting smaller as well because less packets are received by $R$.

Fig.~\ref{fig:sc1d620}-(middle) represents $\bound^c$ and $\bound^r$ and Fig.~\ref{fig:sc1d620}-(bottom) their Pareto-optimal versions $\bound^c_{opt}$ and $\bound^r_{opt}$. Since all packets arriving at $D$ use the $S-R-D$ path, $f_R$ and $f_C$ are equal and thus $\bound^c$ and $\bound^r$ match as expected.
The solutions with a very high $f^c_D$ are the ones where the $S-R$ path is very low and thus, lots of retransmissions would be necessary to overcome the high packet loss probability.
Pareto-optimal bounds ($\bound^c_{opt}$ and $\bound^r_{opt}$) are composed of the solutions concurrently minimizing $f^c_D$ and $f^c_E$. These solutions have a forwarding probability $x_{SR}^{12}=1$ and the relay is located in the very close neighborhood of the center of the $[S,D]$ segment.

For $p_{SD}^1\simeq 0.5$, all bounds are given in Fig.~\ref{fig:sc1d310}. $\bound_{opt}$ is presented in Fig.~\ref{fig:sc1d310}-(top). In $\bound_{opt}$, all solutions of the Pareto set have a perfect link between the relay and the destination (i.e. $p^2_{RD} = 1$).
The lowest capacity ($f_C =0.50$), lowest delay ($f_D=1.00$) and lowest energy ($f_E=1.00$) solution of $\bound_{opt}$ is obtained for solutions where either $p_{SR}^1 = 0$ or $x_{SR}^{12}=0$.
In this case, packets arrive though the direct link $S-D$, minimizing energy and delay since no 2-hop paths are used. The solution with highest capacity ($f_C = 1.50$), delay ($f_D=1.66$) and energy ($f_E=2.00$) has $x_{SR}^{12}=1$ and the relay is in the middle of the $[S,D]$ segment, maximizing $S-R$ and $R-D$ link probabilities ($p_{SR}^1 = 1$ and $p_{RD}^2 = 1$). In this solution, two copies per sent packet are received, one on the direct path, the other on the relay path.
All other solutions from the set $\solSpace$ are as well included in the Pareto solution set $\solSpace_{opt}$. Depending on the relay location and $x^{12}_{SR}$ value, you get either high or low $f_C$. For instance, a solution with a relay close to $D$ and high $x^{12}_{SR}$ has low capacity, delay and energy. A solution with a relay close to $S$ will experience high performance if $x^{12}_{SR}$ is high and low performance if $x^{12}_{SR}$ is low. It shows that the most important variable is the location of the relay, and that the forwarding probability is secondary.

\begin{figure}[h]
\begin{center}
\includegraphics[width=3.3in]{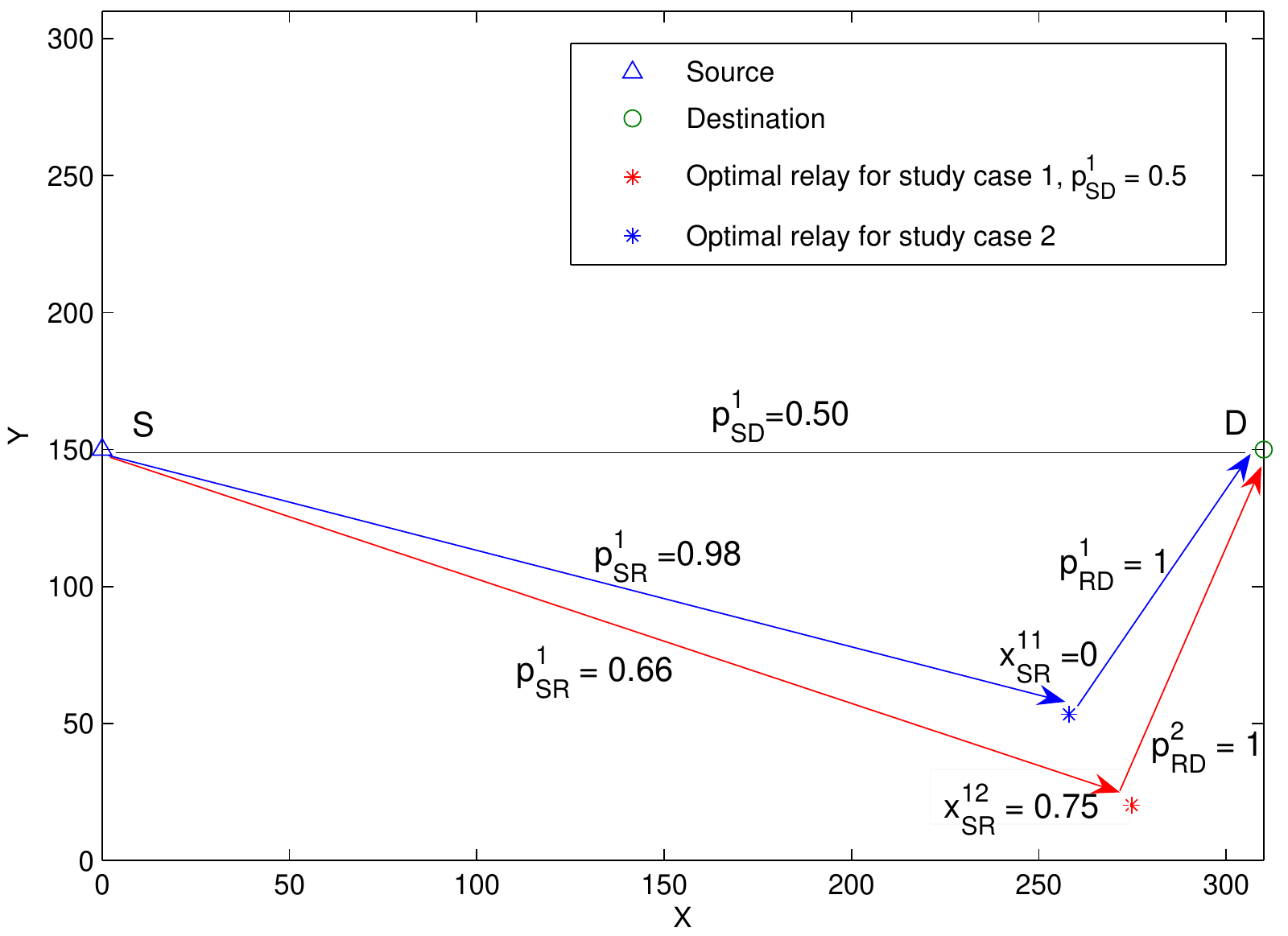}
\includegraphics[width=3.4in]{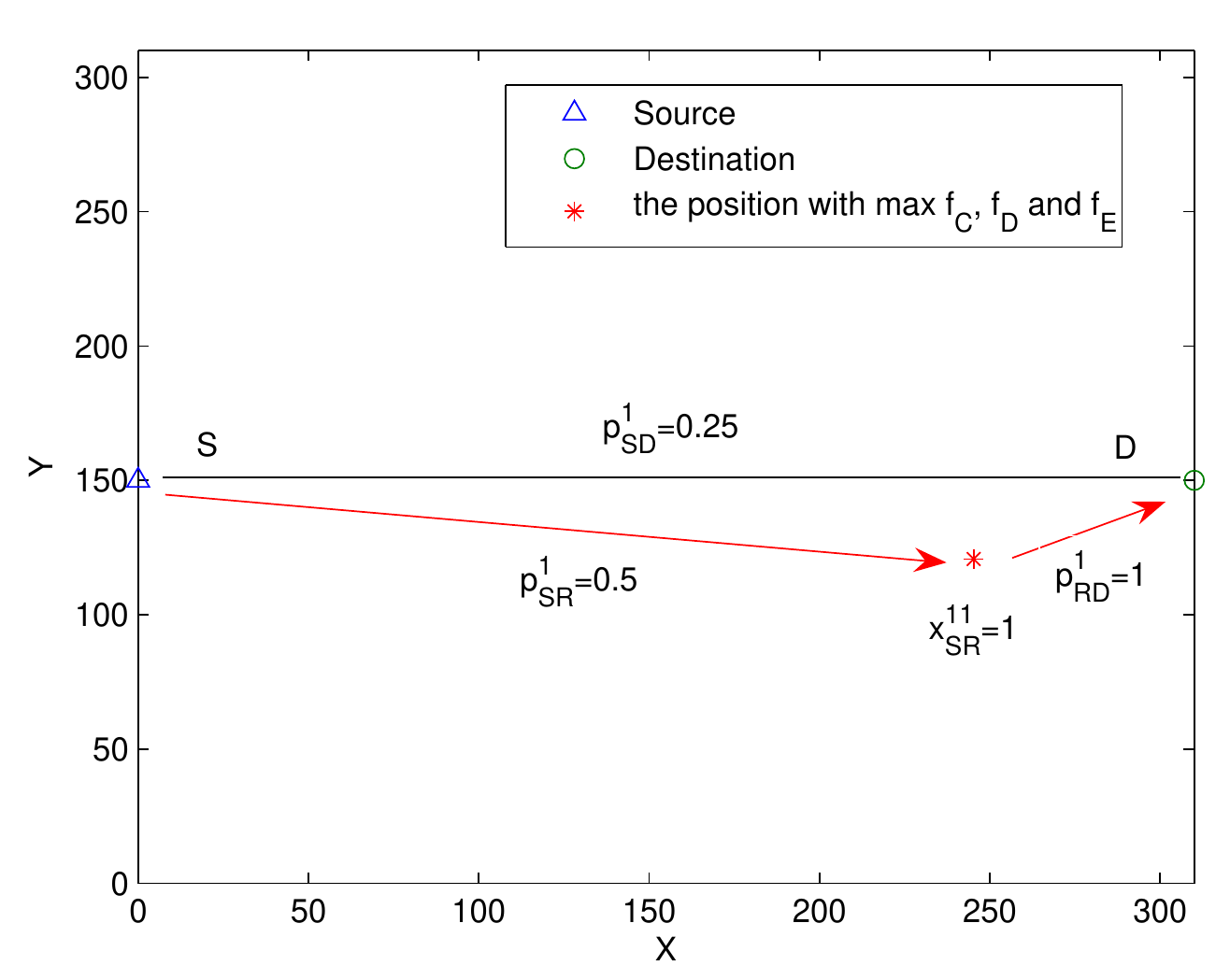}
\caption{Pareto-optimal solutions for study case 1 ($p^1_{SD} \simeq 0.5$) and 2 (top) and solution with maximum $f_C$, $f_D$ and $f_E$ in study case 2, $p^1_{SD}\simeq 0.5$ (bottom). }
\label{fig:optpossc1andsc2}
\end{center}
\end{figure}

Similarly, the capacity-achieving and reliability-achieving bounds are shown in Fig.~\ref{fig:sc1d310}-(bottom), together with $\bound^c_{opt}$ and $\bound^r_{opt}$. In this case, there is  a clear unique Pareto-optimal point with $f^c_D=1.5$ and $f^c_E=1.5$. It is obtained for $f_C = 1$. This Pareto-optimal point contains several solutions.
One of these Pareto-optimal solutions is represented in Fig.\ref{fig:optpossc1andsc2}-(top). All solutions with $f_C=1$ have a perfect $R-D$ link to ensure the forwarded packet perfectly arrives in $D$. Thus, the relay is located closer to $D$ than to $S$. The depicted solution has $x_{SR}^12=1$ and $p_{SR}^1=0.5$. The other solutions have different relay locations and forwarding probability values that verify $p_{SR}^1x_{SR}^{12} = 0.5$ to have $f_C = 1$.

For $p_{SD}^1\simeq 0$, we recall that no packet is transmitted through the $S-D$ link, meaning there are no duplicated packets. Thus $f_C=f_R$ as seen in Fig.~\ref{fig:sc1d620}-(bottom). However, for $p_{SD}^1\simeq 0.5$, the same packet can be transmitted through two paths, creating a difference between reliability and capacity criteria as seen in Fig.~\ref{fig:sc1d310}-(bottom).

\subsubsection{\sc Study Case 2} Transmissions are interference limited.

When $p_{SD}^1\simeq 0$, no relay position ensures $p_{SR}\neq0$ and $p_{RD}\neq 0$ simultaneously, thus no solution exists in this case. It is a direct consequence of interference between $S$ and $R$.

\begin{figure}[t]
\begin{center}
\scalebox{0.7}{
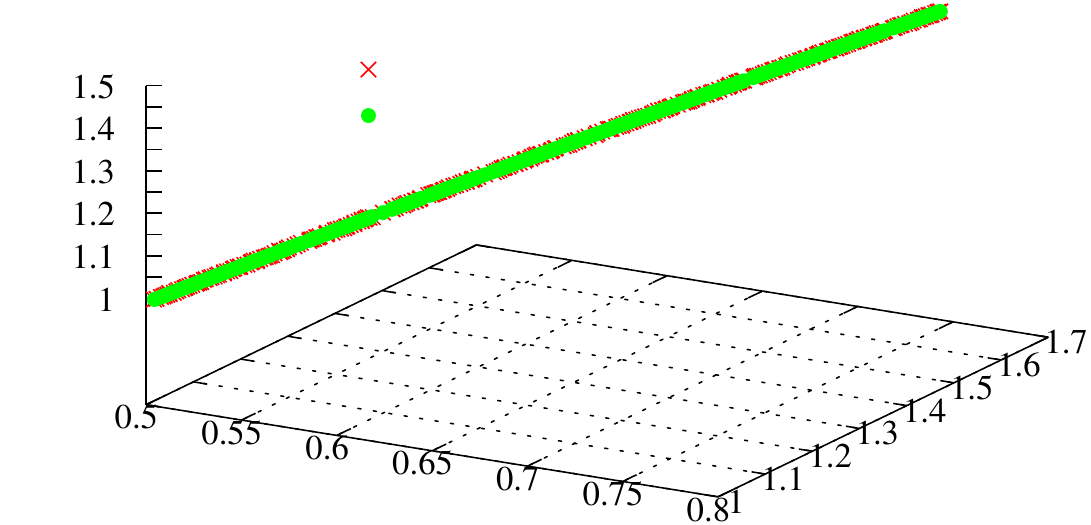
}\vskip8pt
\scalebox{0.6}{
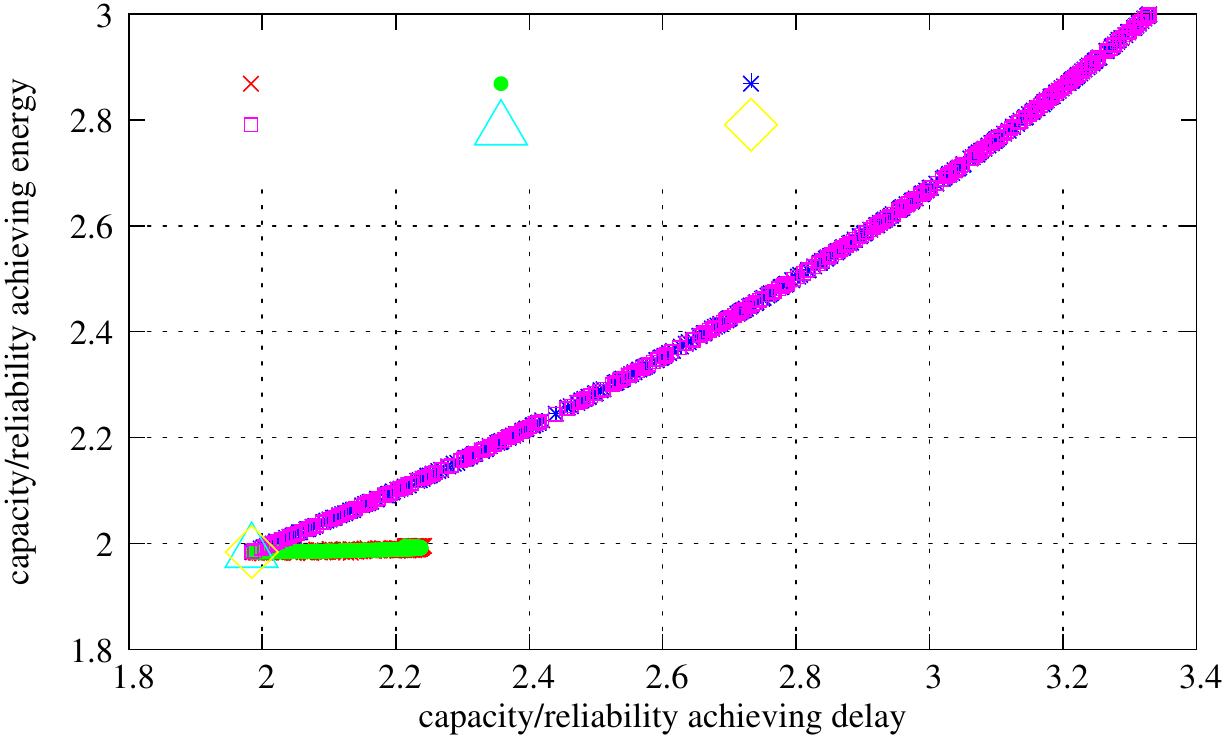
}
\caption{{\sc Study Case 2} with $p_{SD}^1\simeq 0.5$: Pareto-optimal bound $\bound_{opt}$ (top). Capacity and reliability-achieving bounds (bottom).}
\label{fig:sc2d310}
\end{center}
\end{figure}

When $p_{SD}^1\simeq 0.5$, $\bound_{opt}$ is presented in Fig.~\ref{fig:sc2d310}-(top).
For solutions with the lowest capacity ($f_C \simeq 0.5042$), lowest delay ($f_D = 1.0$) and lowest energy ($f_E = 1.0$), the relay doesn't participate in the communication and  packets only arrive in $D$ through the direct link $S-D$.
This is again the case if $R$ is out of reach for $S$ or if $x_{SR}^{11}=0$.
There is only one solution with the highest capacity ($f_C  =0.75$), highest delay ($f_D =1.66$) and highest energy ($f_E = 1.5$). For this solution, the relay always forwards $x_{SR}^{11}=1$ and its location is represented in Fig.~\ref{fig:optpossc1andsc2}-(bottom). Since $R$ and $S$ use the same time slot, interference reduces the maximum link probabilities on links $S-R$ and $R-D$ compared to the no-interference study case 1.
Here, $p_{SR}^1 = 0.5$, $p_{RD}^1 = 1$ and $p_{SD}^1 = 0.25$ and the maximum transmission rate of $R$ is 0.5.
Delay is higher than 1 because some packets arrive on the $S-R-D$ path.

For the solutions different from the minimum and maximum values of the three criteria, $f_C$ decreases for solutions that have a lower forwarding probability. This decrease in $x_{SR}^{11}$ is beneficial to $p^1_{SR}$ and $p^1_{SD}$ since $R$ interferes less with $S$. These solutions have a relay located closer to $D$ to get a perfect link with $D$. Since the relay is forwarding less, capacity, energy and delay decrease.
\begin{figure}[t]
\begin{center}
\scalebox{0.7}{
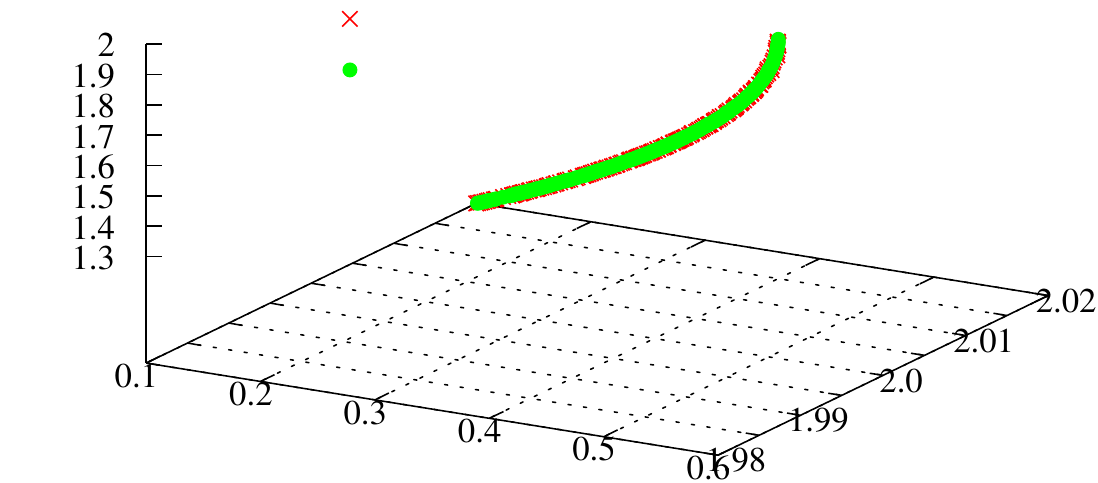
}\vskip7pt
\scalebox{0.6}{
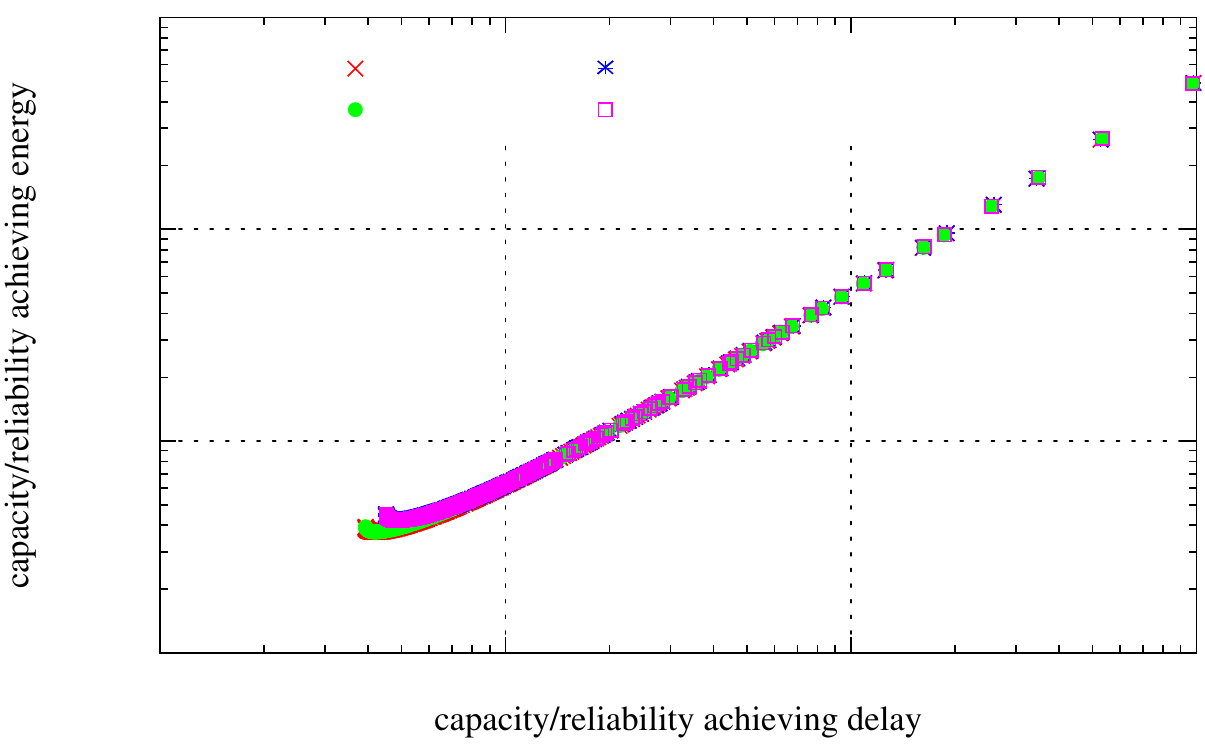
}\vskip7pt
\scalebox{0.6}{
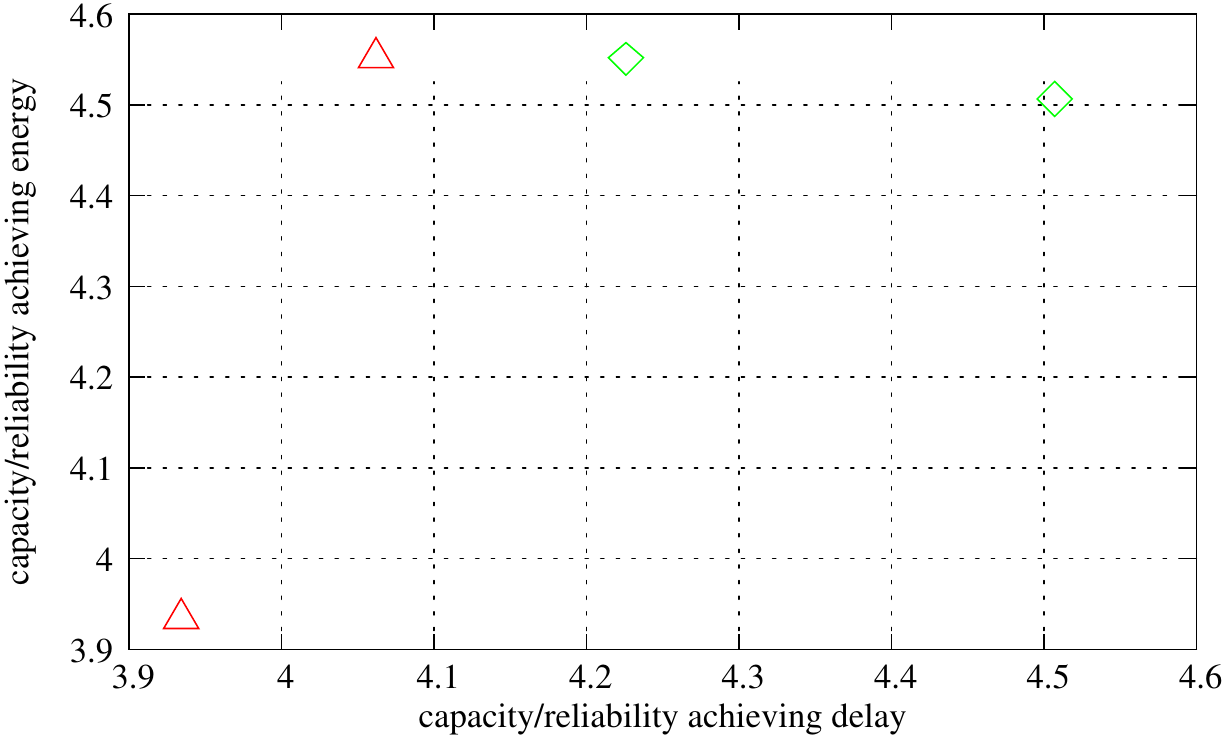
}
\caption{{\sc Study Case 3}: Pareto-optimal bound $\bound_{opt}$ (top). Capacity and reliability-achieving bounds $\bound^c$ and $\bound^r$ (middle) and their corresponding Pareto-optimal bounds  $\bound^c_{opt}$ and $\bound^r_{opt}$ (bottom).}
\label{fig:sc3d620}
\end{center}
\end{figure}
Similarly, the capacity-achieving and reliability-achieving bounds are shown in Fig.~\ref{fig:sc2d310}-(bottom), together with $\bound^c_{opt}$ and $\bound^r_{opt}$.
Since packets may be received from two different path, reliability-achieving and capacity achieving bounds are different. Capacity-achieving represents the upper bound.
Both bounds have the same Pareto-optimal point $\bound^c_{opt}$ and $\bound^r_{opt}$, which is the lowest performance point on $\bound_{opt}$.
This single Pareto-optimal point has $f^c_D=1.98$ and $f^c_E=1.98$ with $x_{SR}^{11} = 0$.
This is the point where the relay is not contributing, thus packets only arrive through the direct path and capacity and reliability coincide.

\subsection{2-relay Pareto bounds and sets} In the 2-relay topology study, all bounds are obtained for $p_{SD}^1 \simeq 0$.

\subsubsection{\sc Study Case 3 } All bounds are given in Fig.~\ref{fig:sc3d620}.
The Pareto optimal bound $\bound_{opt}$ is represented in Fig.~\ref{fig:sc3d620}-(top).
The solution with lowest capacity ($f_C \simeq 0$), lowest delay ($f_D = 2.0$) and lowest energy ($f_e = 1.0$) does not use the relays which are either far away from $S$ or have a null forwarding probability.
The solution with highest capacity ($f_C = 0.508$), highest delay ($f_D = 2.0$) and highest energy ($f_E = 2.0$) is leveraging the two relays. Both relays are located in the middle of $[S-D]$ with $p_{SA}^1 = p_{AD}^2 = 0.504$ (and $p_{SB}^1 = p_{BD}^3 = 0.504$). They use the maximum forwarding probability with $x_{SA}^{12}=1$ and $x_{SB}^{13}=1$.
%For the solutions that are in between minimum and maximum performance, $f_C$ decreases relays are located around the middle of S-D but they are getting close to the D with decrease of $p_{SA}$ and $p_{SB}$ and $x_{SA}^{12}$ and $x_{SB}^{13}$ are decreasing but not too much.

\begin{figure}[h]
\begin{center}
\includegraphics[width=3.4in]{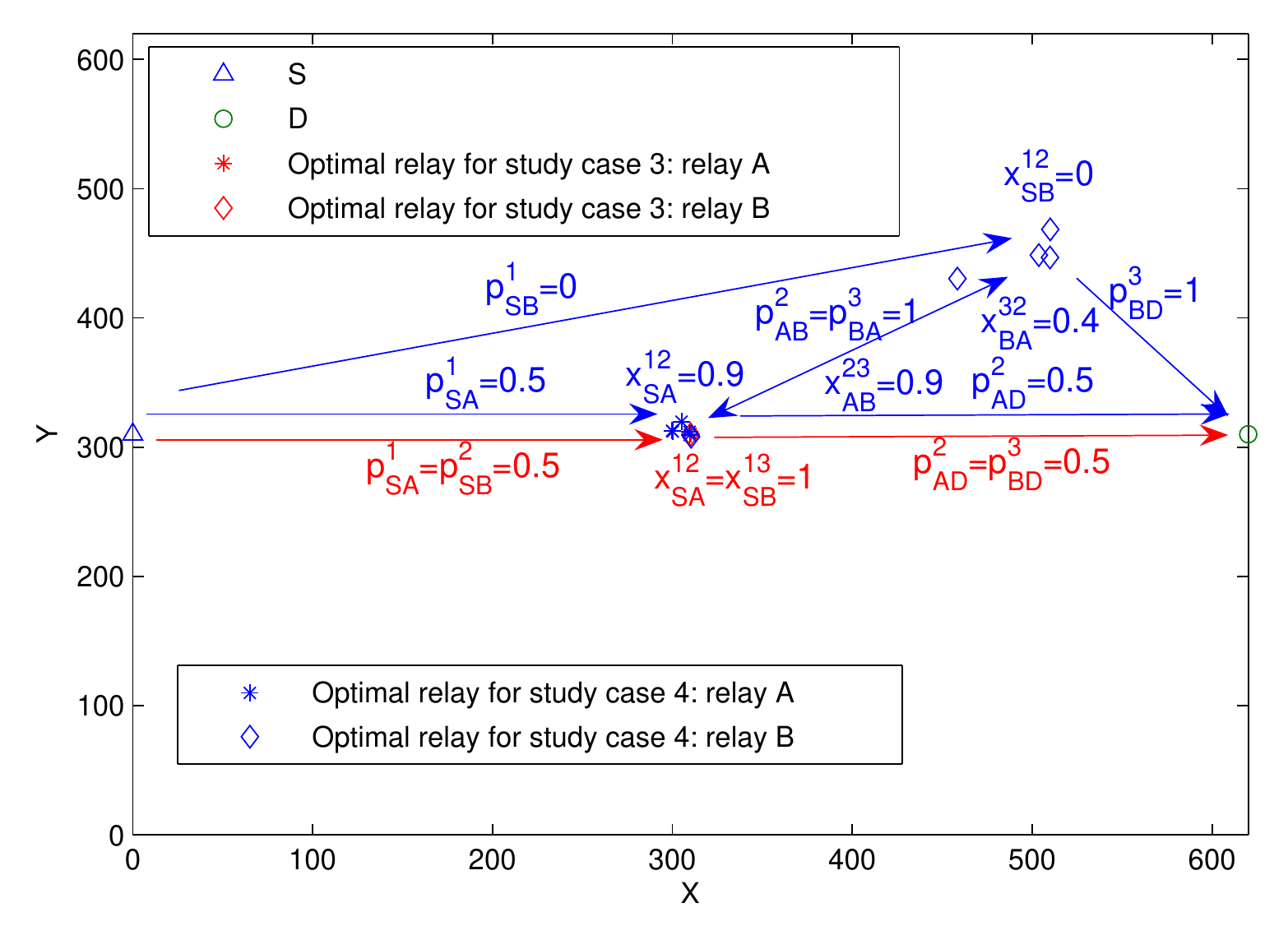}
\caption{Pareto-optimal solutions for study cases 3 and 4}
\label{fig:optpossc3andsc4}
\end{center}
\end{figure}

The capacity-achieving and reliability-achieving bounds are shown in Fig.~\ref{fig:sc3d620}-(middle), and their Pareto-optimal counterpart $\bound^c_{opt}$ and $\bound^r_{opt}$ in Fig.~\ref{fig:sc3d620}-(bottom).
$\bound^c$ and $\bound^r$ don't coincide because packets can arrive from two different paths.
There is a single Pareto-optimal point $f^c_D=3.93$ and $f^c_E=3.93$. It is obtained for $f_C = 0.508351$. This solution has $x_{SA}^{12} = 1$ and $x_{SB}^{13} = 1$ and relays are located exactly in the middle of $[S, D]$ as depicted on Fig.~\ref{fig:optpossc3andsc4}. This solution is the highest capacity case, thus $f^c_D$ and $f^c_E$ are minimized. There is a gap between the upper capacity-achieving bound and the lower reliability achieving bound as represented in Fig.~\ref{fig:sc3d620}-(bottom), due to multi-path transmissions.

\begin{figure}[t]
\begin{center}
\scalebox{0.7}{
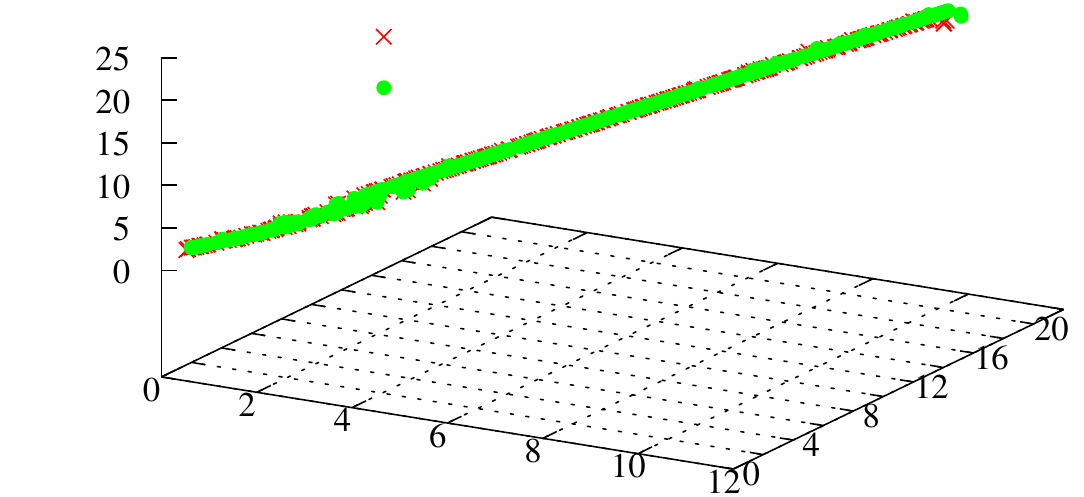
}\vskip7pt
\scalebox{0.6}{
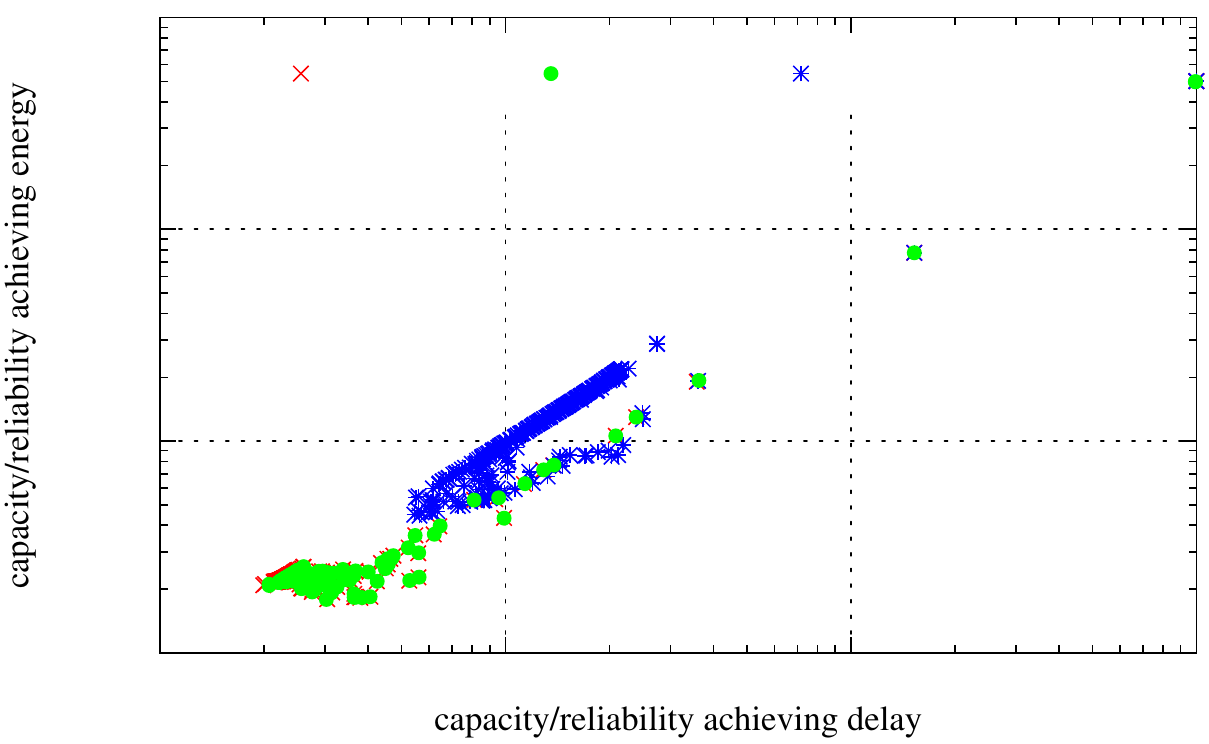
}\vskip7pt
\scalebox{0.6}{
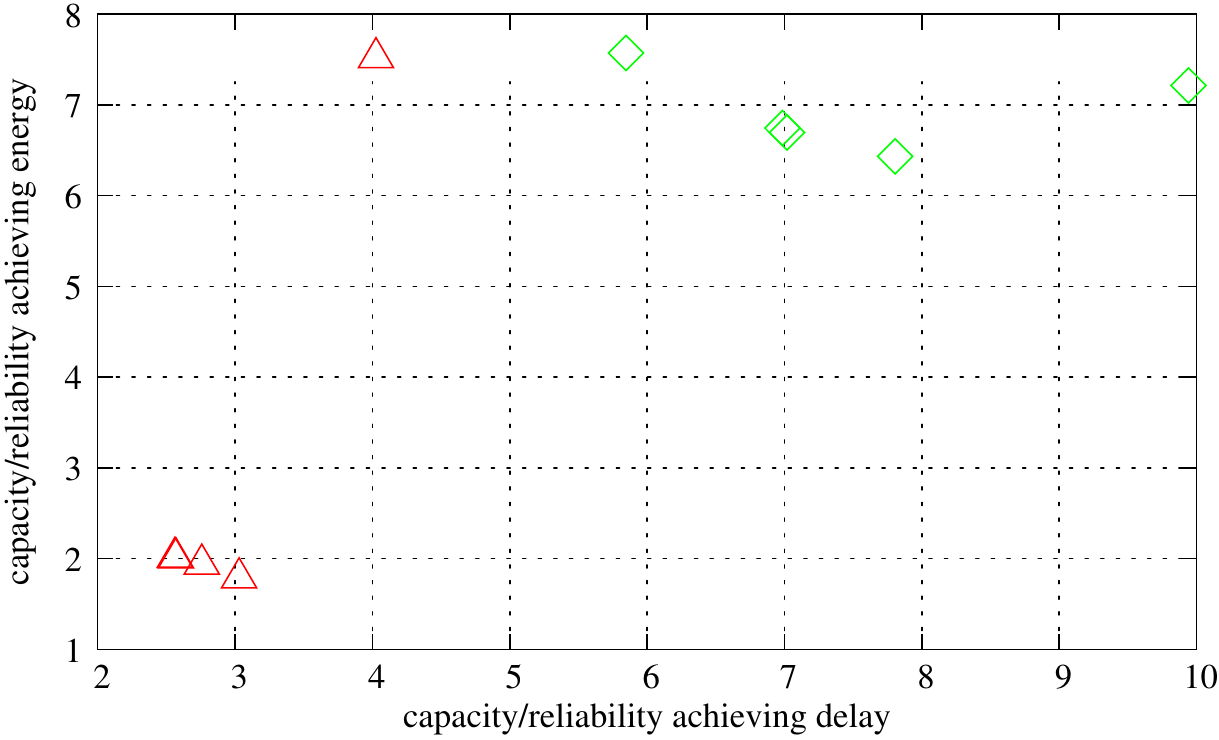
}
\caption{Study Case 4: Pareto-optimal bound $\bound_{opt}$ (top). Capacity and reliability-achieving bounds $\bound^c$ and $\bound^r$ (middle) and their corresponding Pareto-optimal bounds  $\bound^c_{opt}$ and $\bound^r_{opt}$ (bottom).}
\label{fig:sc4d620}
\end{center}
\end{figure}

\subsubsection{\sc Study Case 4 }
The Pareto optimal bound $\bound_{opt}$ is represented in Fig.~\ref{fig:sc4d620}-(top).
Similarly to the previous study cases, the solution with lowest capacity ($f_C \simeq 0$), lowest delay ($f_D = 2.0$) and lowest energy ($f_E = 1.0$) doesn't use the relays.

The solution with the highest capacity ($f_C = 10.16$), highest delay ($f_D =21.00$) and highest energy ($f_e = 21.15$) has its two relays located around the middle of $[S, D]$ with a forwarding probability $x_{SA}^{12}=x_{SB}^{13}=1$. It is in the middle of $[S, D]$ that $p_{SA}^1$ and $p_{AD}^2$ (reps. $p_{SB}^1$ and $p_{BD}^2$) are maximal and equal to 0.5.
There is no interference and the channel between the relays is good with $p_{AB}^{23} =p_{BA}^{32}= 1$.
The difference with study case 3 is that here, packets can be forwarded in the loop between $A$ and $B$. Relays use the maximum forwarding probability between themselves with $x_{SA}^{12}= x_{SB}^{13}= 0.95$.

For the other solutions, one relay ($A$) is located closer to $S$ and the other one ($B$) is closer to $D$. The decrease in $f_C$ is experiences in two ways. Either the relay $A$ is getting further from $S$, reducing $p_{SA}^1$, or  $x_{SA}^{12}$ is decreased and less packets arrive at $A$. Solutions with low values of $f_C$ have small forwarding probability in between relays.

The capacity-achieving and reliability-achieving bounds are shown in Fig.~\ref{fig:sc4d620}-(middle), and their Pareto-optimal counterpart $\bound^c_{opt}$ and $\bound^r_{opt}$ in Fig.~\ref{fig:sc4d620}-(bottom). As expected, $\bound^c$ and $\bound^r$ are disjoint because of the multiple copies received because of the loop between $A$ and $B$.
There are three Pareto-optimal solution in Fig.~\ref{fig:sc4d620}-bottom. Their location and forwarding probabilities are depicted in Fig.~\ref{fig:optpossc3andsc4}.
The optimal solutions are obtained for $f_C \approx 1$. For these solutions, one relay is in the middle of $[S, D]$ and the other relay is close to $D$, having $p^1_{SB} = 0$ and $p^2_{BD} = 1$. The two relays are close together, inducing a perfect link between them with a forwarding probability adjusted to obtain $f_C \approx 1$.

\subsubsection{\sc Study Case 5}
For study cases 5, the two relays share the same time slot and thus transmission is interference-limited.

\begin{figure}[t]
\begin{center}
\scalebox{0.75}{
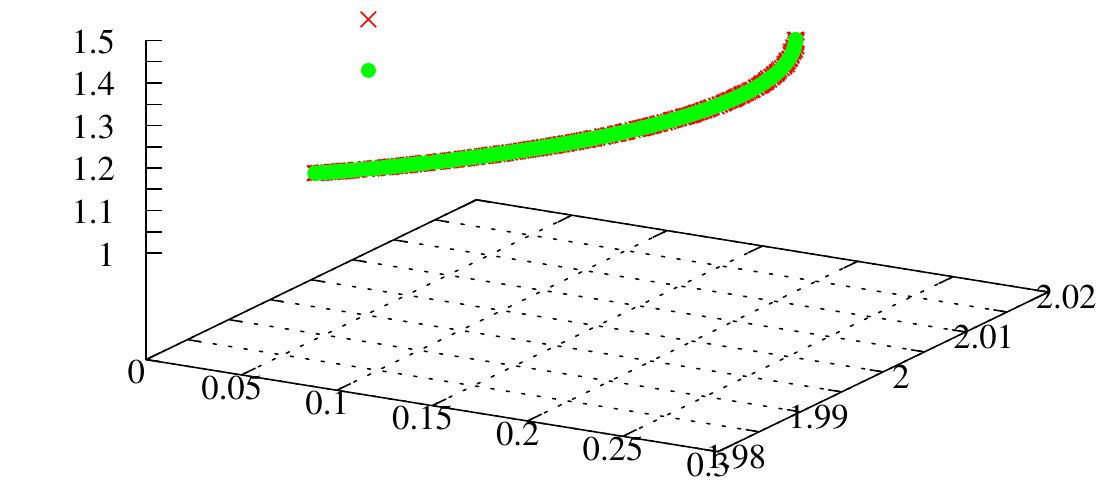
}\vskip7pt
\scalebox{0.6}{
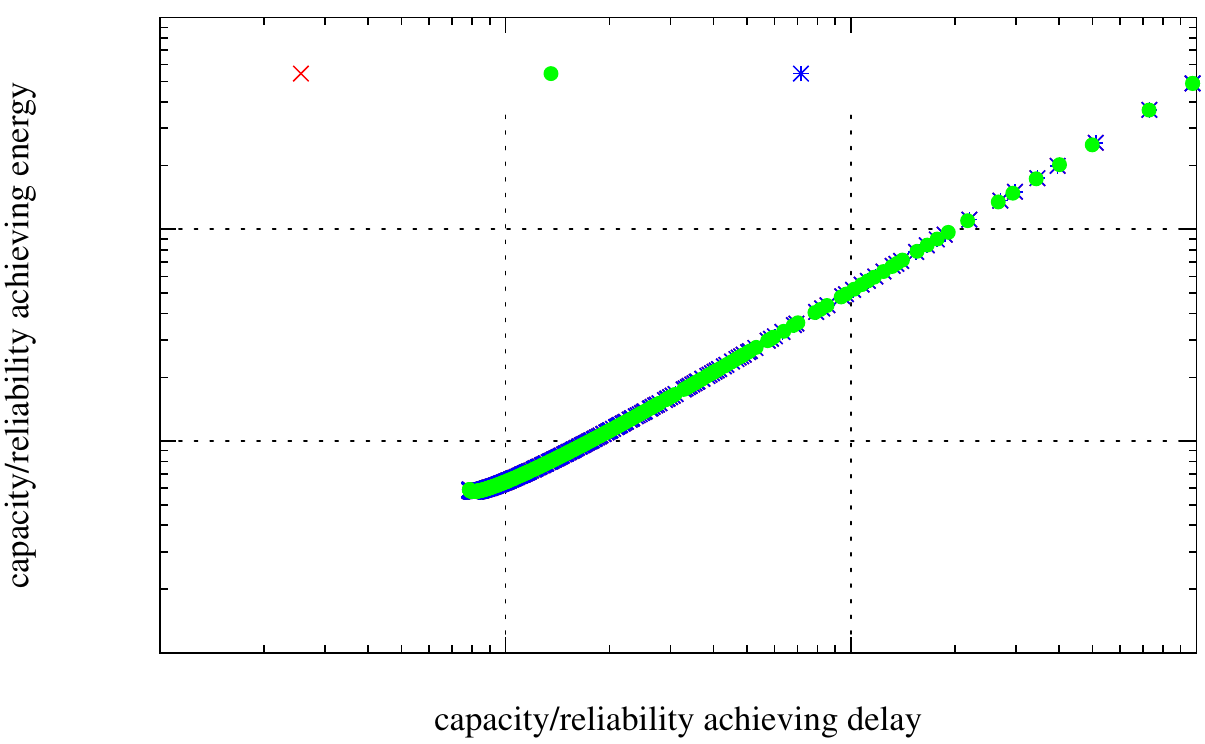
}\vskip7pt
\scalebox{0.6}{
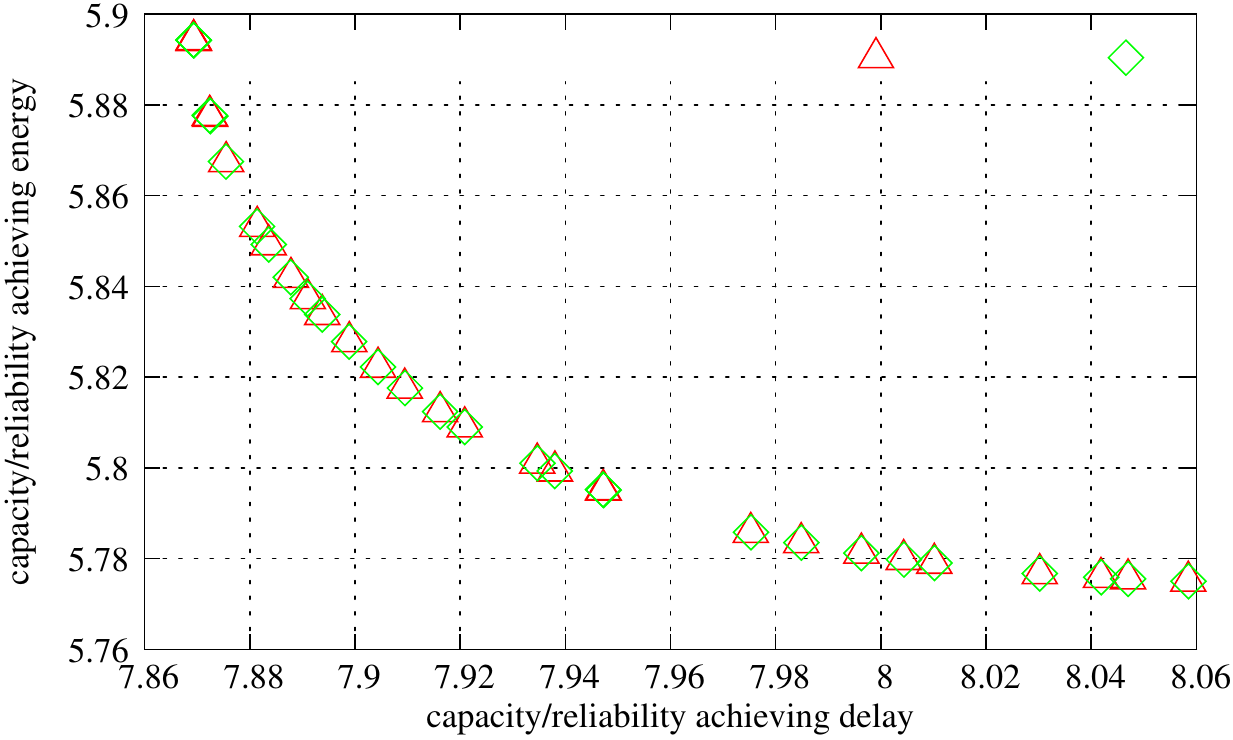
}
\caption{Study Case 5: Pareto-optimal bound $\bound_{opt}$ (top). Capacity and reliability-achieving bounds $\bound^c$ and $\bound^r$ (middle) and their corresponding Pareto-optimal bounds  $\bound^c_{opt}$ and $\bound^r_{opt}$ (bottom).}
\label{fig:sc5d620}
\end{center}
\end{figure}

The Pareto optimal bound $\bound_{opt}$ is represented in Fig.~\ref{fig:sc5d620}-(top).
This bound is the same as the bound observed for study case 1 with $p^1_{SD}\simeq 0$.
For the solution with the highest capacity ($f_C = 0.254$), highest delay ($f_D =2.00$) and highest energy ($f_E = 1.498$), one of its relays is in the middle of $[S, D]$ with a forwarding probability $x^{12}_{SA}=1$ and $p^1_{SA} = 0.5$.
The other relay is not participating in the transmission: it is either located far from $S$ or $x^{12}_{SB} = 0$.
It is the same maximum performance solution than the one observed in Fig.~\ref{fig:sc1d620}-(top) for study case 1.
The same type of observation can be made for the solutions with lowest capacity ($f_C \simeq 0$), lowest delay ($f_D=2$) and lowest energy ($f_E = 1$), where neither relay $A$ nor $B$ are used.

Interference between the relays is clearly detrimental to the network performance since solutions with a single relay dominate solutions with two relays. There are no solutions in $\bound_{opt}$ with two active relays.

\subsection{\sc Comparative analysis}

The purpose of this section is to compare the different Pareto-optimal capacity and reliability achieving bounds.
First, results related to the case where transmission between $S$ and $D$ is possible half the time ($p^1_{SD}=0.5$) is investigated. Second, results related to the case where transmission between $S$ and $D$ is almost impossible ($p^1_{SD}\simeq 0$) are analyzed.

\subsubsection{Case $p^1_{SD}=0.5$}

When one out of two packets can be transmitted on the $S-D$ path, two different 1-relay strategies have been compared. The first one assigns a different time slot to $S$ and $R$ (study case 1) and the other one assigns the same time slot (study case 2). Not surprisingly, the Pareto bounds $\bound^r_{opt}$ and $\bound^c_{opt}$ for the interference free case dominate the one for the interfered scenario.

What is interesting to note, is that for the study case 1, the capacity-achieving upper bound $\bound^c_{opt} = (1.5, 1.5)$ dominates $\bound^r_{opt} = (2, 2)$. If the relay is able to leverage copies to transmit information that hasn't arrived yet through the direct path to $D$, then, average delay could be better than 2 hops but no shorter than 1.5 hops.
Similarly, energy could be smaller than 2 transmissions but no better than 1.5 transmissions.

We will show in the next section \ref{sec:coding} that the combination of source and network coding is the mean to improve the reliability-achieving bound and get closer to the capacity-achieving upper bound.

\subsubsection{Case $p^1_{SD}=0$}

In this case, transmission is almost impossible between $S$ and $D$. We have studied different study cases and we aim at comparing their performance.
First, we have seen that optimizing the problem where the two relays use the same channel in study case 5 converges to a bound where only one relay is active. Thus, interference limited solutions are not surprisingly dominated by interference free solutions. The conclusion is that for bigger networks, optimizing their performance should be done in two steps. First, derive an interference free channel allocation if possible and second, optimize the node's forwarding decisions.

\begin{figure}[h]
\begin{center}
\scalebox{0.7}{
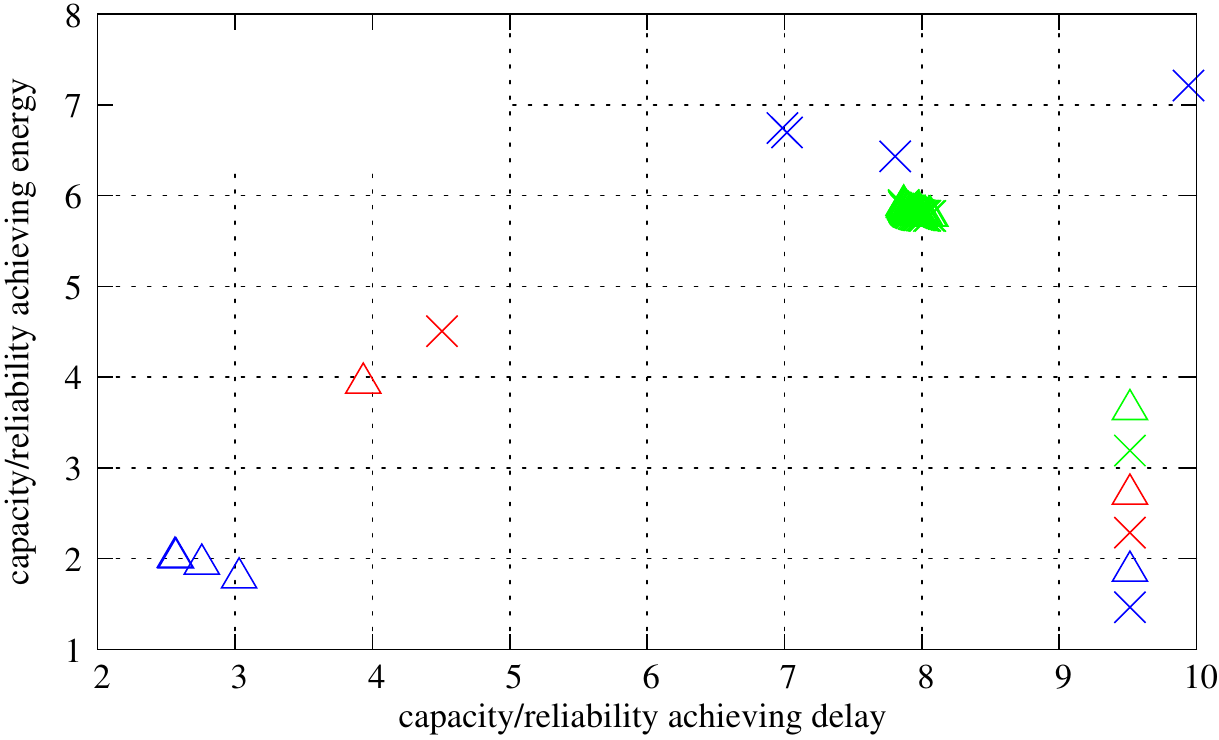
}
\caption{$\bound^c_{opt}$ and $\bound^r_{opt}$ for study cases 1, 3 and 4 where $p_{SD}\simeq0$}
\label{fig:comparison}
\end{center}
\end{figure}

We compare the optimal upper and lower bounds $\bound^c_{opt}$ and $\bound^r_{opt}$ obtained for the interference-free study cases 1, 3 and 4 in Fig.~\ref{fig:comparison}. Looking at the reliability-achieving bound, better performance is obtained if two relays are used since the $\bound^r_{opt}$ bound for study case 3 dominates the bound for study case 1. More reliable transmission are obtained when two relays can be leveraged using our broadcast forwarding mode.

It is really interesting to look at the capacity-achieving bound $\bound^c_{opt}$ for study case 4.
Study case 4 is the only one where a loop exists between $A$ and $B$ in the network.
For this case, lots of copies of the same packet arrive at $D$ because of the loop.
So if it is possible to leverage all these copies using network coding, the network performance can be greatly improved since $\bound^c_{opt}$ dominates the bounds of all other study cases.

If $\bound^c_{opt}$ can be reached, then the optimization of the network forwarding probabilities may be simplified.
There is no need to introduce constraints that avoid the presence of loops in the network. With such a broadcast oriented forwarding mechanism, loops become beneficial for network performance if network coding is used. This is a major contribution of this study. Next, we provide a simple two-layered coding approach and show its benefits for the aforementioned study cases.

\section{Two-layered coding solution to reach $\bound^c_{opt}$ }\label{sec:coding}
In the previous section, we showed that if it is possible to spread the information in the redundant packets forwarded by the relays, it is possible to improve overall network performance. We will show in this section that it is possible to breach the gap between the reliability-achieving lower bound and the capacity-achieving upper bound.
The strategy we propose to leverage the redundant packets in the transmission relies on two design strategies:
\begin{itemize}
\item the use of fountain codes to ensure end-to-end reliability,
\item the use network coding to introduce diversity in the received packets.
\end{itemize}
Introducing coding capabilities requires the introduction of an additional memory of size $M$ which stores the last $M$ packets received by a node.

\begin{figure}[h]
\begin{center}
\includegraphics[width=2.8in]{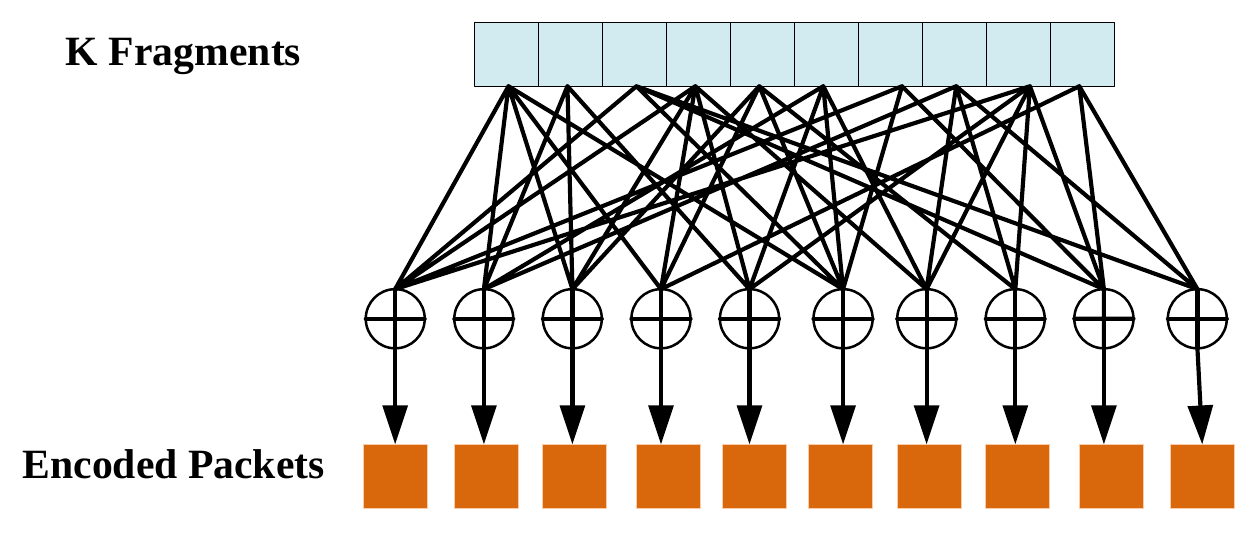}
\caption{Encoding procedure for RL code}
\label{fig:Visio-fountain}
\end{center}
\end{figure}

\subsection{Coding solution}
\subsubsection{Fountain Codes}\label{sec:Fountaincode}
Fountain codes are rateless erasure codes in the sense that a potentially limitless sequence of encoding packets can be generated from the source information. This flow is stopped by the destination when it has received enough packets to recover the information \cite{Mackay2005}. As a consequence, the major advantage of fountain codes is that they are not channel-dependent, thus the same coded information flow is inherently adapted to any channel types. Besides, these codes ensure perfect reliability on the link.

There exists several class of fountain codes. In this paper, we will consider the random linear fountain code(RL code)\cite{Mackay2005}. Indeed, this code requires only 1.6 additionnal overhead packets in average for decoding information with any K fragments. This is an obvious advantage in contrast to Luby Transform code (LT code)\cite{Luby2002}, where the overhead is higher and depends on K. Besides, the RL code is more XOR-friendly, and so better adapted to network coding schemes.
However, the decoding process of RL code is computationally more complex than LT code, since it corresponds to solving a dense linear system of equations. The encoding and decoding computations cost grows as quadratic and cubic respectively with the number of packets encoded, but this scaling is not important if K is less than 1000 \cite{Mackay2005}.

\paragraph*{RL encoding algorithm}
The information from source is first partitioned into K fragments with equal length as shown in Fig.~\ref{fig:Visio-fountain}. Each fragment is selected randomly with probability 1/2 to be XORed to create a new encoded packet. New packets are along those lines created in order to be transmitted until the information can be recovered at destination.

\paragraph*{RL decoding algorithm}

At destination, the received encoded packets are equivalent to equations forming a linear system (where the variables are the fragments). To recover the original information, the system must be full rank. The most efficient decoding algorithm for any random codes on an erasure channel is Maximum Likelihood decoding (ML-decoding), which solves linear equations and can be performed using Gaussian elimination.
%The ML-decoding provides small error probability of decoding, however its complexity can grow rapidly with both N (number of received packets) and K (code length), in the order of O(NK)\cite{Apavatjrut2011}.

\subsubsection{Network coding strategies}\label{sec:codingstrategy}

Network coding is a technique which consists in combining (with XOR operation) packets at the relays.
This introduces packet diversity at the destination as the received packets are more likely to be independent \cite{Apavatjrut2011}.
\begin{algorithm}[h]
\caption{Intra flow network coding algorithm}
\label{alg:XOR}
\begin{algorithmic}
\FOR{each relay node $i$}
\IF{relay node $j$ received a packet $p$ from $i$ at time slot $u$}
\STATE Store the packet $p$ into FIFO memory of size $M$
\IF {$\sum_v x_{ij}^{uv} > 0$}
\STATE $p_{xor}$~= {combine}($p$, FIFO);
\STATE Generate a random value $x\in [0,1]$;
\FOR{($v=1$; $v\leqslant|\T|$; $v=v+1$)}
\IF{$x_{ij}^{uv}\leqslant x$}
\STATE Transmit encoded packet $p_{xor}$ in time slot~$v$;
%\STATE break;
\ENDIF
\ENDFOR
\ENDIF
\ENDIF
\ENDFOR
\end{algorithmic}
\end{algorithm}
Two intra-flow coding strategies that follow Algorithm \ref{alg:XOR} are investigated to take advantage of the multiple copies traveling in the network.
We show that the increase in packet diversity that is created is an efficient mean to distributively reach the theoretical upper MO bound.
In these strategies, when a packet is received, it triggers with a probability $x_{ij}^{uv}$ the emission of the XOR of some packets in the buffer.
The two proposed strategies differ in the way the packets to be XORed are selected.

{\sc Coding Strategy ``R-XOR"}: the XOR operation is made between the lastly R received packets \cite{Apavatjrut2011} as presented in Algorithm \ref{alg:RXOR} considering $M=R$.
\begin{algorithm}[h]
\caption{$p_{new}= \mathrm{combine}(p,MEM)$ }
\label{alg:RXOR}
\begin{algorithmic}
\STATE $p_{new}=p$;
\FOR{each packet $p_{k} \ne p$ in $MEM$}
\STATE $p_{new}= p_{new} \oplus p_{k}$ ;
\ENDFOR
\RETURN $p_{new}$;
\end{algorithmic}
\end{algorithm}

{\sc Coding Strategy ``RLNC"}: A binary Random Linear Network Coding (coding over ${F}_2$) \cite{Ho2006} is performed. For each packet in a relay's buffer, the relay flips a coin to know whether tot add it or not in $p_{out}$, as shown in Algorithm \ref{alg:RLNC}. It is the same as computing an RL code with the packets in memory of the relay. It makes sense to do it since we are sending RL encoded packets at the source. A memory of size of $M=K$ is assumed with $K$ the dimension of the RL code.
\begin{algorithm}[h]
\caption{$p_{new}= \mathrm{combine}(p,memory)$ }
\label{alg:RLNC}
\begin{algorithmic}
\STATE $p_{new}=p$~;
\FOR{each packet $p_{k} \ne p$ in $MEM$}
\STATE Generate a random value $p_{rand}\in [0,1]$;
\IF{($p_{rand}\leqslant 0.5$)}
\STATE $p_{new}=p_{new} \oplus p_{k} $;
\ENDIF
\ENDFOR
\RETURN $p_{new}$~;
\end{algorithmic}
\end{algorithm}

%Ê B- Results
%Ê Ê 1. Performance of coding strategies
%	-> Comparison on the 2-relay no loop case of the different coding strategies : RLNC works the best
%Ê Ê 2. Lower bounds for 2-relay topology
%	-> Give the lower bound obtained only for RLNC and XOR 8 packets for : 2-relay loop case and 2-relay no loop with interference.Ê

\subsection{Lower bounds with coding} \label{sec:results}

\subsubsection{Coding simulations setup}
We consider a message divided into $K$ fragments whose length is the size of a packet. Transmissions are time multiplexed where one packet can be transmitted in one time slot. Note that a frame of $|\T|$ time slots is repeated until the end of simulation. The source sends one RL encoded packet to $D$ in the first time slot of each frame. Location of the relays and their forwarding probability from $\bound_{opt}$ are used. $S$ ends the transmission of RL packets as soon as $D$ can recover the original message and acknowledge the successful reception.

The use of a fountain code at the sources guaranties reliability of the network transmission strategy. Following is the derivation of capacity-achieving delay and energy metrics in this context.
\paragraph*{Capacity achieving delay}
To be consistent with our empirical derivation of the capacity-achieving delay presented in Section \ref{subsec:Implementation}, we derive $\tilde f^c_D$ as following.
We assume that when the coding process ends, the number of packets $N_{TXs}$ that $S$ has transmitted is derived by tracing the last packet that has triggererd the decoding at D.
When coding is used, the capacity-achieving delay is given by:
\[\tilde f^c_D  = \frac{\sum_{h=1}^{h_{max}} h\cdot P(h)}{\frac{K}{N_{TXs}}}
\]
with $P(h)=n(h)/N_{rx}$ the statistical distribution of the delays where $n(h)$ is the number of packets arrived in $h$ hops at $D$ and $h_{max}$ the maximum number of hops of all packets collected at $D$.
Here, $\frac{K}{N_{TXs}}$ is the equivalent of the capacity criterion when coding is used.

%Besides we divide by $K$ to derive the average delay for one sent packet so as to be consistent with the no-coding scheme.
%Then $\tilde f^c_D$ can be derived as
%\[\tilde f^c_D = \frac{N_{TXs}-1+h-1}{K}
%\]
%Here, the reason why we divided by K is that we can derive average delay for one packet so as to be consistent with the no-coding scheme.

\paragraph*{Capacity achieving energy}
The energy consumption is measured by summing the total number of packets transmitted by the source $N_{TXs}$ and the relays $N_{TXr}$ divided by $K$ for normalization. Thus, $\tilde f^c_E$ can be computed as:
\[\tilde f^c_E = \frac{N_{TXs}+\sum_{r=1}^{N}{N_{TXr}}}{K}
\]

%\Katia{Replace with GD definition here and explain that is measures the distance between two bounds.}
The distance between two bounds is measured using the Generational Distance (GD) metric defined as:
\[GD= \frac{1}{N_{opt}}{\left(\sum_{i=1}^{N_{opt}}(d_{i}^{p})\right)^{1/p}}
\]
Where $d_{i}$ is the euclidian distance between the two geometrical points of coordinates $(f^c_D, f^c_E)$ and $(\tilde f^c_D, \tilde f^c_E)$
 \[d_{i} = \sqrt{ (f_{D}^{c}(i)-\tilde f_{D}^{c}(i))^2 + (f_{E}^{c}(i)-\tilde f_{E}^{c}(i))^2 }
 \]
 Here, we use p = 2. The smaller this metric is, the closer the solutions of lower and upper bounds are from each other.
%The distance between two bounds is measured using the Generational Distance (GD) metric defined as
%\[GD= \frac{1}{N_{opt}}{\left(\sum_{i=1}^{N_{opt}}(d_{i}^{p})\right)^{1/p}}
%\]
%Where $d_{i}$ is the euclidian distance between the two geometrical points of coordinates $(f^c_D, f^c_E)$ and $(\tilde f^c_D, \tilde f^c_E)$
%\[d_{i} = \sqrt{ (f^c_{D}(i)-\tilde f^c_{D}(i))^2 + (f^c_{E}^{c}(i)-\tilde f^c_{E}(i))^2 }
% \]

%The error between our model and simulations is measured for each criterion with a normalized root-mean-square error $RMSE= \frac{1}{N_{opt}}\sqrt{\sum_{i=1}^{N}\frac{(f(i)-\tilde f(i))^2}{f(i)^2}}$
%where $N_{opt}$ is the total number of Pareto-optimal solutions in $\bound^c_{opt}$.

%Ê Ê 1. Performance of coding strategies
%	-> Comparison on the 2-relay no loop case of the different coding strategies : RLNC works the best
%Ê Ê 2. Lower bounds for 2-relay topology
%	-> Give the lower bound obtained only for RLNC and XOR 8 packets for : 2-relay loop case and 2-relay no loop with interference.Ê

\subsection{Lower bound results}
%\Katia{Qi, for study case 4 (2-relay, no loop, no interf), do the following simulations
%\begin{itemize}
% 	\item Test RL code only with K=100, 500
%	\item Test RL code K=100, 500 with XOR R=4 and 8
%	\item Test RLNC code K=100, 500
%	\item Create one figure with :  (RL code only K=100, 500) AND (RL code K= 500 with XOR R=4 and 8) : 4 sets
%	\item Create another figure with (RLNC code K=100, 500) AND (RL code K=100, 500 with XOR 8) : 4 sets
%\end{itemize}
%}

In this section, we investigate the performance of coding strategies for the study cases 1 ($p_{SD}^1\simeq 0.5$), 3 and 4 where there is a gap between the upper bound $\bound^c_{opt}$ and the lower bound $\bound^r_{opt}$.
Coding introduces an overhead composed of the additional packets needed to decode the encoded stream.
The overhead is measure using the following criterion:
\begin{equation}
overhead= \frac{E}{K}*100 = \frac{1}{K}{(N_r - K)}*100
\label{eq:overhead}
\end{equation}
where $K$ is the number of initial fragments, $E$ is the number of packets received in excess when using RL codes and $N_{r}$ is the number of packets received at $D$ before decoding the initial fragments.

The transmission of coding coefficients in the encoded packets is an additional overhead.
In this paper, we consider that coefficients are coded over $K$ bits and that the packet length $PL$ is 2560 bytes.
Thus the overhead related to coefficients is given by $(PL- K)/PL$.
This overhead has been taken into account when empirical delay and energy metrics are derived.

\begin{figure}[h]
\begin{center}
\scalebox{0.7}{
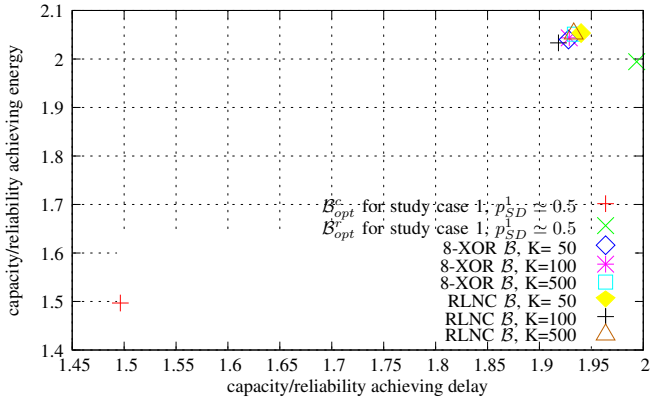
}
\caption{NC lower bounds for 1-relay study case 1 and $p_{SD}^1\simeq 0.5$}
\label{fig:NCboundsc1}
\end{center}
\end{figure}

\subsubsection{{\sc Study case 1} for $p_{SD}^1\simeq 0.5$}
To analyze the performance of the network coding strategies ``R-XOR" and ``RLNC", we set $K$ to 50, 10, and 500 respectively.
Table~\ref{tab:RMSE} presents for each study case and coding strategy the values of the generational distance and the coding overhead of Eq.~\eqref{eq:overhead}.
As shown in Fig.~\ref{fig:NCboundsc1}, the plot shows that coding in this case doesn't improve much the lower bound. This rather negative result can be explained by two reasons. First, since $p_{SD}^1\simeq 0.5$, the reliability is already high ($f_r = 0.75$), reducing the impact of coding. Second, the benefits of coding are here lost by the coding overhead and coefficient transmission. This can be deduced from the energy performance which is slightly worse than for the no coding case.

\begin{figure}[t]
\begin{center}
\scalebox{0.7}{
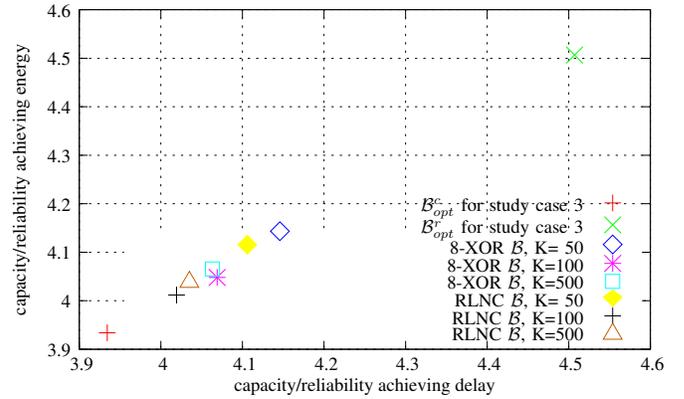
}
\caption{NC Lower bound for 2-relay study case 3.}
\label{fig:NCboundsc3}
\end{center}
\end{figure}
\subsubsection{\sc Study case 3}
Considering $K = 50, K =100$ and $K = 500$ for ``R-XOR" and ``RLNC" respectively, the lower bound results are shown in  Fig.~\ref{fig:NCboundsc3}.
In this scenario, network coding greatly improves the lower bound $\bound^r_{opt}$ and provides bounds that are very close to the capacity-achieving upper bound.

Looking at the impact of $K$ for the ``R-XOR" strategies, it can be seen that with the increase of $K$, the coding bound gets closer to the upper bound. It makes sense since the code dimension increases and the number of overhead packets become more negligible compared to the size of the initial data.

$K$ has not exactly the same impact on the bound for ``RLNC" strategies. This is due to the increase with $K$ of the overhead due to the coefficient stored in the encoded packets.
For $K = 50$ and $K =100$, the coefficients represent 0.24\% and 0.49\% of the packet. Overhead due to coefficients being rather stable, the increase of $K$ is beneficial for the same reasons than for ``R-XOR" strategies. But for $K = 500$, coefficients use 2.46\% of the encoded packet size. This drastic increase is reducing the benefit of using a higher dimension code. Thus, the best ``RLNC" strategy is to use $K=100$.

For the same dimension $K$, ``RLNC" clearly outperforms ``R-XOR" since its lower bound is closer to the upper bound $\bound^c_{opt}$. However, we can note that this improvement is obtained at the cost of bigger buffer at the relays.

\subsubsection{\sc Study case 4}
\begin{figure}[t]
\begin{center}
\scalebox{0.7}{
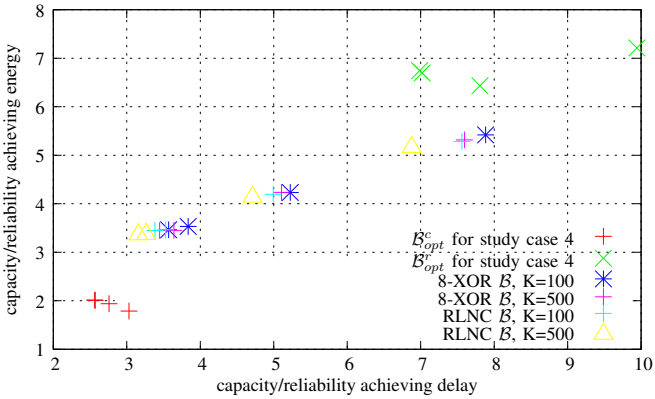
}
\caption{NC Lower bound for 2-relay study case 4.}
\label{fig:NCboundsc4}
\end{center}
\end{figure}

Here, the bounds for $K =100$ and $K = 500$ for ``R-XOR" and ``RLNC" strategies are derived on Fig.~\ref{fig:NCboundsc4}.
Different from the study case 3,  the increase of $K$ is improving the coding lower bounds. Here, the code dimension has a positive impact on the higher number of copies received at the relay. We recall that in study case 4, a loop exists. Thus, the number of overhead packets in the coding solution is really smaller than the number of redundant packets in    the no-coding lower bound.

Similarly to the study case 3, ``RLNC" outperforms ``R-XOR".
The main result of this paper is that we have exhibited a coding strategy that provides a performance bound that is really close to the capacity-achieving upper bound $\bound^c_{opt}$.
We can conclude that the capacity-achieving bound we have defined in this paper is a very tight bound on the multiobjective performance of the network.
The simple source and network coding strategies presented in the paper are efficient for study cases where we can leverage path diversity. Studying the capacity-achieving bound is an efficient mean to characterize the Pareto-optimal performance with respect to delay and energy consumption for a network using a broadcast forwarding paradigm.

To better understand the coding impact for these coding strategies, we further look into the overhead compared to the ideal RL coding.
In the ideal situation for RL code, the number of excess packets is equal to 1.611970 in our simulation environment.
Thus, the overhead proportion for the ideal situation is equal to 3.2239\%, 1.6120\% and 0.3224\% for K = 50, 100 and 500 respectively.
The closer the overhead to the ideal RL coding is, the better the network coding strategy is efficient. The generational distance and overhead are shown in Table~\ref{tab:RMSE}.

Seen from this table, the best coding strategy for study case 1 ($p_{SD}^1\simeq 0.5$) is when adopting the RLNC strategy for K = 100 with the lowest value of generational distance equal to 0.6863.
For the study case 3, the best coding strategy is also adopting the RLNC strategy for K = 100 with the lowest value of generational distance equal to 0.1152.
However, for the study case 4, the best coding strategy is when adopting the RLNC strategy for K = 500 with the lowest value of generational distance equal to 1.57.
This means that RLNC strategy gives results very close to the optimal theoretical bound. But considering the transmission of coding coefficients, the performance
doesn't always increase with K.

\begin{table}
\caption{The RMSE and overhead for different coding strategies}
\centering
\label{tab:RMSE}
\begin{tabular}{|l|l|l|l|} \hline
%\multicolumn{2}{|c|}{Study Cases} & GD & Overhead (\%) \\ \hline
Study Cases & Coding Strategies& GD & Overhead (\%) \\ \hline
\multirow{6}{*}{\minitab[c]{Study case 1\\($p_{SD}^1\simeq 0.5$)}} & 8-XOR, K=50 & 0.6901 & 43.06 \\
& 8-XOR, K=100 & 0.6908& 41.782 \\
& 8-XOR, K=500 & 0.7061& 40.562 \\
& RLNC, K=50 & 0.7118& 42.98 \\
& RLNC, K=100 & 0.6863& 41.094 \\
& RLNC, K=500 & 0.7065& 40.383\\ \hline
\multirow{6}{*}{Study case 3} & 8-XOR, K=50 & 0.298 & 5 \\
& 8-XOR, K=100 & 0.1764 & 2.31 \\
& 8-XOR, K=500 & 0.1843 & 0.784 \\
& RLNC, K=50 & 0.25 & 4.34 \\
& RLNC, K=100 & 0.1152 & 2.05 \\
& RLNC, K=500 & 0.1455 & 0.394\\ \hline
\multirow{4}{*}{Study case 4} & 8-XOR, K=100 & 1.8561 & 57.62 \\
& 8-XOR, K=500 & 1.7704 & 55.68 \\
& RLNC, K=100 & 1.7431 & 56.215 \\
&RLNC, K=500 & 1.57 & 55.664\\ \hline
\end{tabular}
\end{table}

\section{Conclusion}\label{sec:conclusion}\label{sec:conclu}
This paper has presented a flexible framework for evaluating the performance of simple wireless relay networks with respect to several performance criteria. It has been designed to account for the broadcast nature of wireless communications and for an accurate interference characterization for the network. This framework allows for the determination of two lower and upper Pareto bounds and their corresponding Pareto solutions. Network model and bounds for 1-relay and 2-relay networks have been assessed though simulations. We have shown that the upper MO bound provides a tight bound on the performance of network coding strategies. Thus, this work not only confirms the accuracy of our optimal theoretical bound, but also proposes a way of approaching it as close as wanted.
This work will be extended to tackle problems where more relays belong to the network of interest. The problem will as well be formulated for the case where several concurrent flows transit in the network.
\appendix
\section{Appendix}\label{sec:Appendix}
This Appendix details the derivation of $f_C$ in Eq.~\eqref{eq:fc}, $f_D$  Eq.~\eqref{eq:fd} and $f_E$ in Eq.~\eqref{eq:fe} for the 2-relay cases.
\subsection{Capacity criterion $f_C$}
$f_C$ is defined as the average number of packets received by the destination per packet sent by $S$. It is  derived by adding the success probabilities of a packet arriving at $D$ through all possible path as defined in Eq.~\eqref{eq:capacite}.
For example, for the direct path $S-D$, the success probability equals $\tau_S^1p_{SD}^1$. For the relay path $S-A-D$, the capacity equals to $\tau_S^1Q_{SA}^{12}p_{AD}^{2}$. Similarly, the success probability for other paths can be derived as shown in Table ~\ref{tab:notations1}.
\begin{table}
\caption{Path analysis for capacity criterion}
\centering
\label{tab:notations1}
\begin{tabular}{| c | l |}
\hline
Path	        & Path success probability \\\hline
S-D		& $\tau_S^1p_{SD}^1$ \\\hline
S-A-D		& $\tau_S^1Q_{SA}^{12}p_{AD}^{2}$ \\\hline
S-B-D		& $\tau_S^1Q_{SB}^{13}p_{BD}^{3}$ \\\hline
S-A-B-A-D		& $\tau_S^1Q_{SA}^{12}(Q_{AB}^{23}Q_{BA}^{32})p_{BD}^{3}$ \\\hline
S-B-A-B-D		& $\tau_S^1Q_{SB}^{13}(Q_{AB}^{23}Q_{BA}^{32})p_{BD}^{3}$ \\\hline
S-A-B-D		& $\tau_S^1Q_{SA}^{12}Q_{AB}^{23}p_{BD}^{3}$ \\\hline
S-A-B-A-B-D		& $\tau_S^1Q_{SA}^{12}(Q_{AB}^{23}Q_{BA}^{32})^2p_{BD}^{3}$ \\\hline
S-B-A-D		& $\tau_S^1Q_{SB}^{13}Q_{BA}^{32}p_{AD}^{2}$\\\hline
S-B-A-B-A-D		& $\tau_S^1Q_{SB}^{13}(Q_{BA}^{32}Q_{AB}^{23})^2p_{AD}^{2}$ \\\hline
\vdots             & \vdots \\
\hline
\end{tabular}
\end{table}

%\begin{equation}
%\begin{array}{cc}
%f = f_{SD} + f_{SAD} + f_{SBD} + f_{SABAD} + f_{SBABD}+ \\
%f_{SABD} + f_{SABABD} + f_{SBAD} + f_{SBABAD} + ...
%\end{array}
%\end{equation}
The sum of the success probabilities for all paths is the sum of the terms of the following infinite geometric series:
\begin{equation}
\begin{split}
f_C  =   \tau_S^1p_{SD}^{1} + (E + F)(1 + Q_{AB}^{23}Q_{BA}^{32} + (Q_{AB}^{23}Q_{BA}^{32})^2 +\\
+ Q_{AB}^{23}Q_{BA}^{32})^3 +\cdots + (Q_{AB}^{23}Q_{BA}^{32})^n)  \nonumber
%\tau_S^1 [p_{SD}^1 + \frac{1}{} \\ \left[Q_{SA}^{12}(p_{AD}^2+Q_{AB}^{23}p_{BD}^3)+Q_{SB}^{13}(p_{BD}^3+Q_{BA}^{32}p_{AD}^2) \right]]
\end{split}
\end{equation}
with $E = (\Qe{S}{A}{1}{2} + \Qe{S}{B}{1}{3}\Qe{B}{A}{3}{2})p_{AD}^{2}$ and $F = (\Qe{S}{B}{1}{3} + \Qe{S}{A}{1}{2}\Qe{A}{B}{2}{3})p_{BD}^{3}$. Here, $\tau_S^1(E + F)$ is the first term of the series, and $Q_{AB}^{23}Q_{BA}^{32}$ is the common ratio.
As $n$ goes to infinity, the absolute value of $Q_{AB}^{23}Q_{BA}^{32}$ must be less than one for the series to converge. This is true since we add the constraint $Q_{AB}^{23}\leq 1-\Delta$ and $Q_{BA}^{32}\leq 1-\Delta$ ($\Delta = 0.05$) in our MO problem. The sum then becomes:
\begin{equation}
\begin{split}
f_C  =   \tau_S^1p_{SD}^{1} + \frac{\tau_S^1}{1-Q_{AB}^{23}Q_{BA}^{32}}(E + F)  \nonumber
%\tau_S^1 [p_{SD}^1 + \frac{1}{} \\ \left[Q_{SA}^{12}(p_{AD}^2+Q_{AB}^{23}p_{BD}^3)+Q_{SB}^{13}(p_{BD}^3+Q_{BA}^{32}p_{AD}^2) \right]]
\end{split}
\end{equation}

\subsection{Delay criterion $f_D$}
\begin{table}
\caption{Path analysis for delay criterion}
\centering
\label{tab:notations2}
\begin{tabular}{| c | l |}
\hline
Path	        & Delay per path \\\hline
S-D		& $f_{SD} = p_{SD}^1$ \\\hline
S-A-D		& $f_{SAD} = 2Q_{SA}^{12}p_{AD}^2$ \\\hline
S-B-D		& $f_{SBD} = 2Q_{SB}^{13}p_{BD}^3$ \\\hline
S-A-B-A-D		& $f_{SABAD} = 4Q_{SA}^{12}(Q_{AB}^{23}.Q_{BA}^{32})p_{AD}^2$ \\\hline
S-B-A-B-D		& $f_{SBABD} = 4Q_{SB}^{13}(Q_{BA}^{32}.Q_{AB}^{23})p_{BD}^3$ \\\hline
S-A-B-D		& $f_{SABD} = 3Q_{SA}^{12}Q_{AB}^{23}p_{BD}^3$ \\\hline
S-A-B-A-B-D		& $f_{SABABD} = 5Q_{SA}^{12}Q_{AB}^{23}(Q_{BA}^{32}.Q_{AB}^{23})p_{BD}^3$ \\\hline
S-B-A-D		& $f_{SBAD} = 3Q_{SB}^{13}Q_{BA}^{32}p_{AD}^2$\\\hline
S-B-A-B-A-D		& $f_{SBABAD} = 5Q_{SB}^{13}Q_{BA}^{32}(Q_{AB}^{23}Q_{BA}^{32})p_{AD}^2$ \\\hline
\vdots             & \vdots \\
\hline
\end{tabular}
\end{table}

$f_D$ is defined as the average delay a packet sent by the source needs to reach the destination. It is calculated by summing the delays for all packets arriving through all possible paths and dividing the result by the number of copies $f_C$ as defined in Eq.~\eqref{eq:delay}.
A similar path analysis is done for the delay computation in Table~\ref{tab:notations2}.
For example, for the direct path $S-D$, the packet arrives in $D$ in one hop and the corresponding delay equals $f_{SD} = p_{SD}^1$. For the relay path $S-A-D$, the packet takes two hops to arrive at $D$ and thus the delay of the path equals to $f_{SAD} = 2Q_{SA}^{12}p_{AD}^2$.
The infinite sum of the delays of Table~\ref{tab:notations2} is performed and provides a delay criterion of:
\begin{equation}
\begin{split}
%f_D  = \frac{1}{f_C} \cdot \frac{\tau_S^1}{(1-Q_{AB}^{23}Q_{BA}^{32})^2}(A+B)
f_D  = \frac{1}{f_C} \cdot [\frac{\tau_S^1}{(1-Q_{AB}^{23}Q_{BA}^{32})^2}(A+B) + \tau_S^1p_{SD}^1]  \nonumber
\end{split}
\end{equation}
with $A = p_{AD}^2[\Qe{S}{B}{1}{3}\Qe{B}{A}{3}{2}(3-\Qe{A}{B}{2}{3}\Qe{B}{A}{3}{2})+2\Qe{S}{A}{1}{2}]$ and $B =p_{BD}^{3}[\Qe{S}{A}{1}{2}\Qe{A}{B}{2}{3}(3-\Qe{A}{B}{2}{3}\Qe{B}{A}{3}{2})+2\Qe{S}{B}{1}{3}]$. Again it originates from the summation of the terms of an infinite series. 

\subsection{Energy criterion $f_E$}
\begin{table}
\caption{Path analysis for energy by relays}
\centering
\label{tab:notations3}
\begin{tabular}{| c | l |}
\hline
Path	        & Energy per path \\\hline
S-D		& $f_{SD} = 0$ \\\hline
S-A-D		& $f_{SAD} = \tau_S^1Q_{SA}^{12}$ \\\hline
S-B-D		& $f_{SBD} = \tau_S^1Q_{SB}^{13}$ \\\hline
S-A-B-A-D		& $f_{SABAD} = \tau_S^1Q_{SA}^{12}(Q_{AB}^{23}Q_{BA}^{32})$ \\\hline
S-B-A-B-D		& $f_{SBABD} = \tau_S^1Q_{SB}^{13}(Q_{BA}^{32}Q_{AB}^{23})$ \\\hline
S-A-B-D		& $f_{SABD} = \tau_S^1Q_{SA}^{12}Q_{AB}^{23}$ \\\hline
S-A-B-A-B-D		& $f_{SABABD} = \tau_S^1Q_{SA}^{12}Q_{AB}^{23}(Q_{BA}^{32}.Q_{AB}^{23})$ \\\hline
S-B-A-D		& $f_{SBAD} = \tau_S^1Q_{SB}^{13}Q_{BA}^{32}$\\\hline
S-B-A-B-A-D		& $f_{SBABAD} = \tau_S^1Q_{SB}^{13}Q_{BA}^{32}(Q_{AB}^{23}Q_{BA}^{32})$ \\\hline
.\vdots             & \vdots \\
\hline
\end{tabular}
\end{table}

$f_E$ is defined as the average number of emissions done by all nodes per packet sent. It is derived by summing the probability for a relays to emit a packet per path and the probability for the source to emit a packet (which is equal to its rate $\tau_S^1$). %Considering the energy by relays, for the direct path(S-D), the energy by relays equals to $f_{SD} = 0$. Meanwhile, for the relay path(S-A-D), the energy by relays equal to $f_{SAD} = \tau_S^1Q_{SA}^{12}$. 
Similarly, the energy consumed per paths is shown in Table ~\ref{tab:notations3} for all possible paths. 
Again, the summation of the terms of an infinite geometric series leads to the following criterion:
\begin{equation}
\begin{split}
f_E =\tau_S^1 + \frac{\tau_S^1}{1-Q_{AB}^{23}Q_{BA}^{32}}(\Qe{S}{A}{1}{2} +  \Qe{S}{B}{1}{3}\Qe{B}{A}{3}{2} \\+ \Qe{S}{A}{1}{2}\Qe{A}{B}{2}{3} + \Qe{S}{B}{1}{3})   \nonumber
\end{split}
\end{equation}

%\bibliographystyle{unsrt}
%\bibliography{IEEEabrv,basebiblio}

\bibliographystyle{IEEEtran}
\bibliography{IEEEabrv,basebiblio}

%\bibliographystyle{plain}
%\bibliography{basebiblio}

\end{document}